\documentclass[aps,prb,twocolumn,epsfig,showpacs,amsmath,float]{revtex4-1}
\usepackage{graphicx,epsf,natbib,amsmath,latexsym,amssymb}
\usepackage{subfigure}
\usepackage{color}
\def\be{\begin{equation}}
\def\ee{\end{equation}}
\def\bea{\begin{eqnarray}}
\def\eea{\end{eqnarray}}
\def\bw{\begin{widetext}}
\def\ew{\end{widetext}}
\def\nn{\nonumber}
\begin{document}
\author{Wan-Ju Li, Sung-Po Chao, Ting-Kuo Lee}
\affiliation{Institute of Physics, Academia Sinica, Taipei 11529, Taiwan}
\title{Controversy over large proximity induced s-wave-like pairing from a d-wave superconductor}

\date{\today}

\begin{abstract}
We use the proximity effect to generate effective topological superconductors
by placing metals with strong spin-orbit coupling in contact with a
superconductor, aiming to produce Majorana zero modes useful for
topologically-protected quantum computation. In recent experiments, several
quintuple layers of $\text{Bi}_2\text{Se}_3$ were epitaxially grown on the
high-$T_c$ material $\text{Bi}_2\text{Sr}_2\text{CaCu}_2\text{O}_{8+\delta}$, and conflicting
experimental results were reported. We use the standard mean-field approach to
study this heterostructure and find it is unlikely to have a large
proximity-induced superconducting gap. Despite the seemingly correct
temperature dependence, the $s$-wave gap claimed to be observed may not be
purely superconducting in origin. Future work on the proximity-induced bulk
superconducting gap and the interfacial bandstructure should shed light on this issue.

\end{abstract}
\maketitle

\section{Introduction}
The proximity effect in superconductor-metal interfaces, or the leakage of
Cooper pairs from the superconductor into the metal, has been studied since its
experimental discovery in 1960\cite{Meissner1960}. The magnitude of the induced superconducting gap generally depends on the quality of the interface, external fields, and certain material-dependent properties, in particular the Fermi surfaces and energy spectra of both the superconductor and metal\cite{Steiner2006,Cirillo2005,Tesauro2005}. The symmetry of the induced superconducting gap is typically identical to that of the superconductor unless there exists symmetry-breaking perturbations, such as the lattice-symmetry mismatch, ferromagnetism in the metal\cite{Bergeret2001,Keizer2006}, or the presence of large spin-orbit interactions\cite{Fu2008}. 

Recently, the application of the proximity effect to the surface states of a topological insulator (TI), with the goal to produce the topological superconductivity, has stimulated intense theoretical and experimental effort. In addition to $s$-wave superconductors\cite{Fu2008,Kim2014,Black-Schaffer2013,Takane2014}, $d$-wave superconductors also show interesting proximity effect on the surface states of a TI\cite{Linder2010, Linder2010a,Parisa2012,Yilmaz2014,Xu2014,Wang2013,Black-Schaffer2013}. Wang et al.\cite{Wang2013} observe a surprisingly large $s$-wave-like pairing gap on the top surface of a seven-quintuple-layer-thick $\text{Bi}_2\text{Se}_3$ film on $\text{Bi}_2\text{Sr}_2\text{CaCu}_2\text{O}_{8+\delta}$ ($\text{BSCCO}$) by using angle-resolved photoemission spectroscopy(ARPES). By applying a suitable incident photon energy (approximately $50$ eV), at which the contribution from the surface states dominates that from the bulk states, a gap of up to $15$ meV on the outermost surface states of the $\text{Bi}_2\text{Se}_3/\text{BSCCO}$ system is detected. The change of the pairing symmetry from $d$-wave in BSCCO to $s$-wave-like on the surface states from the proximity effect is claimed to reflect a nontrivial coupling in this system\cite{Wang2013}. However, two separate experimental groups\cite{Yilmaz2014,Xu2014} with similar setups do not reproduce the proximity-induced pairing gap on the $\text{Bi}_2\text{Se}_3/\text{BSCCO}$. Earlier tunneling measurements\cite{Parisa2012} suggest the proximity induced gap at the interface is $d$-wave.  These conflicting observations prompt us to address the issue theoretically and ask whether such a large proximity-induced s-wave gap is possible.

In this paper, we use the tunneling model pioneered by McMillan\cite{McMillan1968} to study the superconducting proximity effect between $\text{Bi}_2\text{Se}_3$ and BSCCO. We assume the interface between the superconductor and the metal is clean or with only weak random impurities, and the superconducting gap in BSCCO is identical\footnote{Here we treat the BSCCO as the bulk, and its superconductivity is not influenced by the contact surface with $\text{Bi}_2\text{Se}_3$. More rigorously this interface gap should be evaluated self consistently, and the gap magnitude of BSCCO at the interface would be smaller than its bulk value. Considering this would further decrease the proximity induced gap at the $\text{Bi}_2\text{Se}_3$. Thus the gap obtained in the main text should be viewed as the theoretical upper bound.} for all coupling strengths between BSCCO and $\text{Bi}_2\text{Se}_3$. We model the
superconductivity within BSCCO via mean-field, and evaluate within $\text{Bi}_2\text{Se}_3$ the induced pairing amplitude, defined as the sum of the expectation value of Cooper pairs in momentum space, for a range of symmetries. Our results show that the proximity-induced superconducting gap is small and mainly $d$-wave, and the unambiguous way to find the condensates is through the evaluation of pairing amplitudes. The smallness of $s$-wave and $p\pm ip$ pairing amplitudes supports the absence of a large induced $s$-wave-like superconducting gap. 

 Based on our results, we suggest the discrepancies between the experiments\cite{Wang2013,Yilmaz2014,Xu2014} are due to different interface coupling strengths between $\text{Bi}_2\text{Se}_3$ and BSCCO across the various samples. The observed gap in Ref.\onlinecite{Wang2013} may not purely be due to superconductivity, but also to a mass gap generated by bands crossings caused by large interfacial tunnelings as compared with the samples used in Ref.\onlinecite{Yilmaz2014} and \onlinecite{Xu2014}. The growth temperature in Ref.\onlinecite{Yilmaz2014} is certainly much higher than that in Ref.\onlinecite{Wang2013}, suggesting the interface tunneling strength is larger in sample of Ref.\onlinecite{Wang2013}. Further analysis on the symmetry of the bulk superconducting gap of $\text{Bi}_2\text{Se}_3$ in the samples of Ref.\onlinecite{Wang2013} should reveal its $d$-wave nature, as the bands overlap between the bulk bands of $\text{Bi}_2\text{Se}_3$ and BSCCO should be smaller than the overlap between the surface states. Another way to verify our claim is to measure the band structure of $\text{Bi}_2\text{Se}_3$ at the interface, which should show a $d$-wave single particle gap structure with lobes pointing in the nodal direction when the system temperature is higher than $T_c$, the superconducting transition temperature of BSCCO.  

We note that earlier work by Z. X. Li et. al.\cite{Yao2015} has also addressed the same issue with a two-bands model of the $\text{Bi}_2\text{Se}_3$. In their work there are two quantum
well states other than the surface state present at the Fermi surface for the
7QL-thick $\text{Bi}_2\text{Se}_3$ . They also introduce small (around $100$ meV) repulsive and attractive interactions within the $\text{Bi}_2\text{Se}_3$. At the top layer (away form the interface with BSCCO) they find $s$-wave dominant pairing due to the suppression of $d$-wave pairing near the $\Gamma$ point and claim that disorder could further enhance $s$-wave pairing. The proximity induced superconducting gap magnitude in their paper\cite{Yao2015} is also around $1$ meV with similar tunneling strength. Their results support the reports in Ref.\onlinecite{Yilmaz2014,Xu2014} of the absence of a superconudcting gap despite its $s$-wave gap structure. Furthermore, the dominant $s$-wave pairing gap in Ref.\onlinecite{Yao2015} is the result of multilayer structure of $\text{Bi}_2\text{Se}_3$(through enforcing $s$-wave interactions in $\text{Bi}_2\text{Se}_3$ and disorder, which prefer $s$-wave) and choice of Fermi surface around $\Gamma$ point. This mechanism is different from our work where the formation of $s$-wave-like gap is the combined effects of superconducting proximity effect and hybridization of bands between $\text{Bi}_2\text{Se}_3$ and BSCCO.

We organize this paper as follows: In Section.\ref{TIHTC}, we introduce our model Hamiltonians for the BSCCO/$\text{Bi}_2\text{Se}_3$ system. Assuming good contact surface between the two, we compute various paring amplitudes and proximity-induced gaps in the surface state of $\text{Bi}_2\text{Se}_3$ at different tunneling strengths. In Section.\ref{secwarpe} we discuss other possible factors which could enhance the proximity induced superconductivity. In Section.\ref{s wave}, we find the combined effect of strong hybridization of bands and superconducting proximity effect gives rise to $s$-wave-like gap structure. In Section.\ref{discon} we summarize our results and suggest possible experimental approaches to verify our claim.     

\section{Proximity effect: BSCCO and $\text{Bi}_2\text{Se}_3$}\label{TIHTC}
We start with the model Hamiltonian describing the $\text{Bi}_2\text{Se}_3$, a type of three dimensional topological insulator(TI), grown on top of the BSCCO. We assume only single metallic layer in the $\text{Bi}_2\text{Se}_3$ in our computation, which can be viewed as one of the two dimensional surfaces of the three dimensional topological insulator $\text{Bi}_2\text{Se}_3$ in contact with BSCCO if the bulk band were insulating. The motivations for forming this heterostructure\cite{Fu2008,Wang2013,Xu2014,Yilmaz2014} is to produce effective $p\pm ip$ superconductivity, a form of topological superconductivity, on its interface. The edge modes or vertex states of the topological superconductor host the Majorana modes\cite{Fu2008}, the simplest anyon which could be useful for quantum computations. 

To describe the superconducting proximity effect between surfaces of BSCCO and $\text{Bi}_2\text{Se}_3$, we use the following single band bilayer model Hamiltonian:
\bea\label{model Hamiltonian}
H&=&H_{BSCCO}+H_{\text{Bi}_2\text{Se}_3}+H_t\\\nn
H_{BSCCO}&=&\sum_{\vec{k},\sigma}d^{\dagger}_{\vec{k},\sigma}(E_d(\vec{k})-\mu_d)d_{\vec{k},\sigma}\\\nn&+&\sum_{\vec{k},\vec{k'}}V(\vec{k},\vec{k'})d_{\vec{k},\uparrow}d_{-\vec{k},\downarrow}d_{-\vec{k'},\downarrow}^{\dagger}d_{\vec{k'},\uparrow}^{\dagger}\\\nn
H_{\text{Bi}_2\text{Se}_3}&=&\sum_{\vec{k},\alpha,\beta}c^{\dagger}_{\vec{k},\alpha}(\hat{E}_c(\vec{k})_{\alpha\beta}-\mu_c \hat{I}_{\alpha\beta})c_{\vec{k},\beta}\\\nn
H_t&=&\sum_{\vec{k},\sigma}\big(t_{\vec{k}}c^{\dagger}_{\vec{k},\sigma}d_{\vec{k},\sigma}+t_{\vec{k}}^{\ast}d^{\dagger}_{\vec{k},\sigma}c_{\vec{k},\sigma}\big).
\eea
Here $H_{BSCCO}$, $H_{\text{Bi}_2\text{Se}_3}$, and $H_t$ describe the Hamiltonian for two dimensional surface of BSCCO, surface state of $\text{Bi}_2\text{Se}_3$, and the single particle tunneling terms between the two surfaces. H$\sigma$, $\alpha$, $\beta$ denote spin indices, $d_{\vec{k},\sigma}$ and $c_{\vec{k},\sigma}$ the electron annihilation operators of BSCCO and $\text{Bi}_2\text{Se}_3$ in momentum space representation, and $\vec{k}=(k_x,k_y)$ is the linear momentum of the two dimensional surface. $\hat{I}_{\alpha\beta}$ is the identity matrix, and $\hat{E}_c(\vec{k})_{\alpha\beta}$ is a $2\times2$ matrix characterizing the spin orbit interaction of surface state of TI. Attractive $V(\vec{k},\vec{k'})$ with $d$ wave symmetry is assumed to give $d$-wave superconductivity of BSCCO under the BCS mean field approximation. The tunneling amplitude $t_k$ is proportional to the wavefunctions overlap between that of BSCCO and $\text{Bi}_2\text{Se}_3$. We have assumed few or weak random nonmagnetic impurities such that spin and momentum are conserved in the tunneling term, and constant tunneling amplitude nearby the relevant Fermi level ($t_k=t$).

\begin{figure}[h]
\centering
\subfigure[~$E_c(k_x,0)$ vs $k_x$]{
\includegraphics[width=.45\columnwidth]{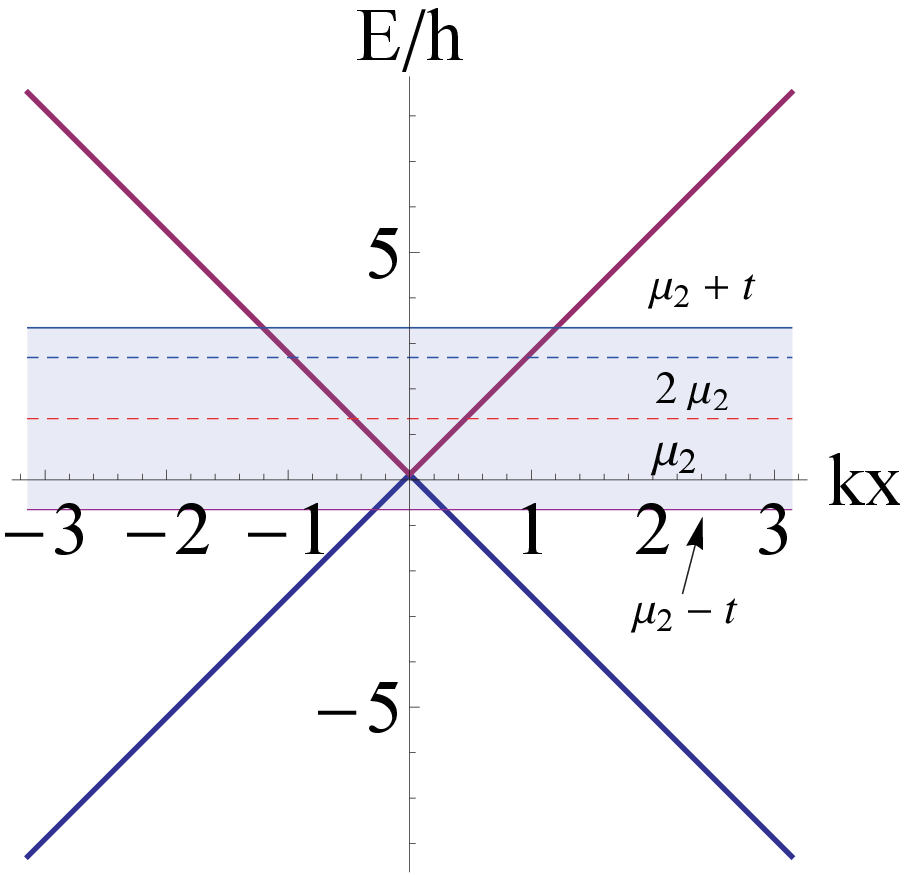}
\label{sFSlat0} }
\subfigure[~FS with warping]{
\includegraphics[width=.45\columnwidth]{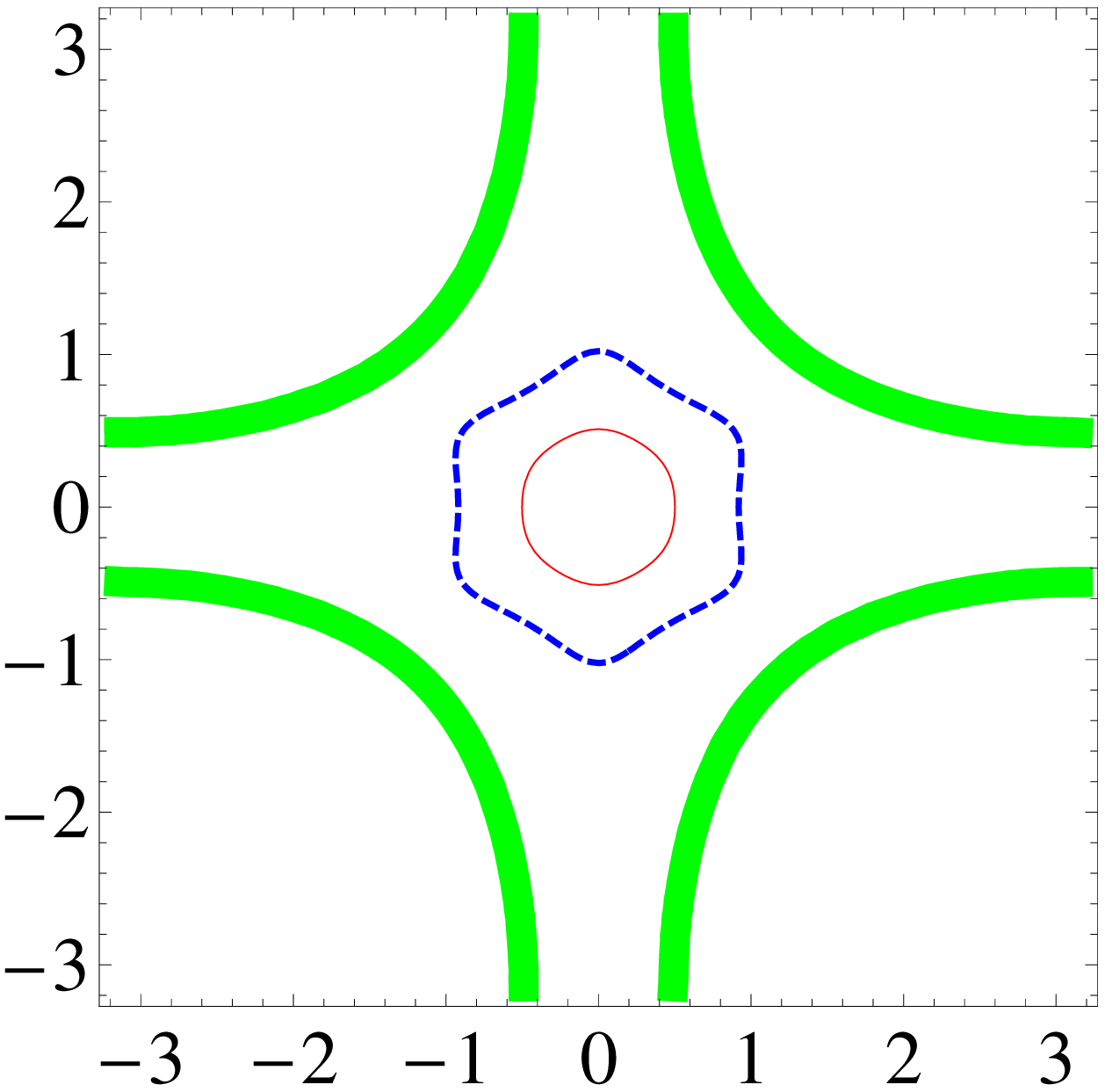}
\label{sFSlat1} }
\caption{(a) Energy dispersion of surface state of $\text{Bi}_2\text{Se}_3$ described in Eq.(\ref{warp2}). The relevant energy scale in the main text is colored in shaded blue region. (b) We plot the Fermi surfaces of $\text{Bi}_2\text{Se}_3$ including the warping cubic terms at $\mu_2/h\simeq1.3$ (red thin line) and $\mu_2/h\simeq2.6$ (blue thick dashed line), and BSCCO (green thick line). The $\mu_2/h\simeq2.6$ is just for illustrating the hexagonal symmetry of the Fermi surface which is difficult to see with the Fermi level ($\mu_2/h\simeq1.3$) set by the experiments.}
\label{sFSlatt0}
\end{figure}

The explicit form of $E_d(\vec{k})$ and $E_c(\vec{k})_{\alpha\beta}$ are obtained from the tight binding models, with parameters specified from first principle studies for $\text{Bi}_2\text{Se}_3$\cite{Liu2010,Zhang2010} and BSCCO\cite{Norman1995}. The single band dispersion for BSCCO is:
\bea\nn
&&\bar{E_d}(\vec{k})\equiv E_d(\vec{k})-\mu_d=-h\left(\cos(k_x a_1)+\cos(k_y a_1)\right)\\\nn&+&h'\cos(k_x a_1)\cos(k_y a_1)-h^{''}(\cos(2k_x a_1)+\cos(2k_y a_1))\\\nn&-& h^{'''}(\cos(2k_x a_1)\cos(k_y a_1)+\cos(k_x a_1)\cos(2k_y a_1))\\\label{bcco}&+&h^{''''}(\cos(2k_x a_1)\cos(2k_y a_1))-\mu_d
\eea
Here $h=0.2975$eV, $h'=0.1636$eV, $h^{''}=0.0259$eV, $h^{'''}=0.0558$eV, $h^{''''}=0.0510$eV, $\mu_d=-0.1305$eV, $a_1\simeq 3.82\AA$. For the $\text{Bi}_2\text{Se}_3$ crystal, to account for the $C_6$ crystal symmetry and different orientations on the proximity effect seen in the experiments, we include the warping terms\cite{Liu2010} in our model Hamiltonian. From Ref.\onlinecite{Liu2010}, the low energy dispersion of $\text{Bi}_2\text{Se}_3$ close to the Brillouin zone center is given by
\bea\nn
\bar{E_c}(\vec{k})&=&E_c(\vec{k})-\mu_c\\\nn&=&p_0+p_1(k_x^2+k_y^2)+p_2(\sigma_x k_y-\sigma_y k_x)\\\label{warp}
&+&p_3(k_+^3+k_-^3)\sigma_z-\mu_c
\eea
where $p_0=0.035$ eV, $p_1=1.38556$ eV$\AA^2$, $p_2=0.795$ eV$\AA$, $p_3=0.3535$ eV$\AA^3$, $k_{\pm}=k_x\pm ik_y$, and $\sigma_x$, $\sigma_y$, and $\sigma_z$ are the $2\times 2$ Pauli spin matrices. The $p_2$ term is the Rashba spin orbit term. For our calculations, we need to extend this low energy Hamiltonian to the whole Brillouin zone in momentum space and keep the single Dirac-cone-like structure. The quadratic $k$ terms have undesired behavior in the large-$k$ region and does not give the consistent band structures compared with experiments at large momenta. To include the warping terms and avoid the issues of inconsistency at large $k$, we keep the linear term unchanged, remove the quadratic terms, and replace $k_x a_2$ with $\sin(k_x a_2)$ and $k_y a_2$ with $\sin(k_y a_2)$ for the third order warping terms. Here $a_2=4.138\AA$ is the lattice constant of $\text{Bi}_2\text{Se}_3$. The low energy dispersion in Eq.(\ref{warp}) is then modified as:
\bea\label{warp2}
\bar{E_c}(\vec{k})&=&p_0+p_2(\sigma_x k_y-\sigma_y k_x)\\\nn
&+&\frac{p_3}{a_2^3}\left(2\sin^3(k_x a_2)-6\sin(k_x a_2) \sin^2(k_y a_2)\right)\sigma_z-\mu_c
\eea
 The chemical potential $\mu_c$ exhibiting surface states of $\text{Bi}_2\text{Se}_3$ is estimated to be around $0.4$eV for the relevant experiment\cite{Xu2014}. Noting that Eq.(\ref{warp2}) is not of tight binding form but from an effective low energy Hamiltonian of surface state of $\text{Bi}_2\text{Se}_3$ with parameters fixed by first principle studies\cite{Liu2010,Zhang2010}. The first order derivative of $\bar{E_c}(\vec{k})$ is discontinuous at zone boundary ($k_x a_2$, $k_y a_2$ being $\pm \pi$) after imposing periodic boundary condition. For the low energy limit (energy scale less than $p_2 a_2\sim 3.2eV$) discussed in this paper this artifact does not enter into our equations or modify our results. The region of relevant energy scale in our discussion here is plotted in Fig.\ref{sFSlat0}.

For the numerical computations shown below we take nearest neighbor hopping $h$ of BSCCO in Eq.(\ref{bcco}) as the energy unit. We use $k_x a_1\equiv \tilde{k_x} $ and $k_y a_1\equiv \tilde{k_y}$ as dimensionless momentum space parameters although there is a minor difference in their lattice size ($a_2/a_1\simeq 1.08$). In this unit the $d$-wave superconducting gap is defined as
\bea\nn
\Delta(\vec{k})^{\ast}&=&\sum_{\vec{k'}}V(\vec{k}-\vec{k'})\langle d_{\vec{k'},\uparrow}^{\dagger}d_{-\vec{k'},\downarrow}^{\dagger}\rangle\\
&=&\Delta^{\ast}(\cos(\tilde{k_x})-\cos(\tilde{k_y}))
\eea
Here the gap magnitude $|\Delta|\simeq 40$ meV (or $|\Delta|/h\simeq 0.13$) is assumed to be fixed (termed the "bulk limit" below) by viewing the superconductivity of BSCCO not influenced by the contact of $\text{Bi}_2\text{Se}_3$. The expectation value is taken with the ground state of the whole mean field Hamiltonian $H$. In principle we may also compute this gap magnitude self consistently, as shown for the model calculations in Appendices.\ref{AA}-\ref{AD}, by fixing the interaction strength $V(\vec{k},\vec{k'})$. The obtained proximity induced superconducting gaps or pairing amplitudes are smaller compared with those obtained with the bulk limit. To obtain an upper bound on the proximity induced gap on $\text{Bi}_2\text{Se}_3$ we stick with this bulk limit in the main text of the paper. 

We identify the proximity induced superconductivity in $\text{Bi}_2\text{Se}_3$ by computing two quantities: the pairing amplitude and the quasiparticle energy gap. The pairing amplitude $A_i$ is defined as:
\bea\label{eq6}
A_i^{\ast}=-\sum_{\vec{k'}}f_i f_i'\langle c_{\vec{k'},\alpha}^{\dagger}c_{-\vec{k'},\beta}^{\dagger}\rangle=\tilde{A_i}f_i
\eea
with $f_i$ being the symmetry factor and $i$ denoting $s$, $p$, $p\pm ip$, or $d$ wave symmetry ($\alpha$ and $\beta$ are spin indices. $f_s=1$, $f_p=\sin(\tilde{k_x})$ or $\sin(\tilde{k_y})$, $f_{p\pm ip}=\sin(\tilde{k_x})\pm i \sin(\tilde{k_y})$, and $f_d=\cos(\tilde{k_x})-\cos(\tilde{k_y})$. $f_i'$ has similar definition as $f_i$ with $\tilde{\vec{k}}$ replaced by $\tilde{\vec{k'}}$. This $f_if_i'$ geometric factor comes from the angular expansions of $V(\vec{k}-\vec{k'})$.) and $\tilde{A_i}$ is the corresponding magnitude. Even (spatial) wave symmetry has odd spin angular momentum and vice versa, as required by the Pauli exclusion principle. This dimensionless pairing amplitude directly reflects the amount of Cooper pairs formed at $\text{Bi}_2\text{Se}_3$, but it is not directly probed by tunneling measurement\cite{Parisa2012} nor spectroscopy like ARPES\cite{Yilmaz2014,Xu2014,Wang2013}. The physical quantity probed in these experiments is the quasiparticle energy gap. We identify the energy gap due to proximity effect in two approaches: One is from the density of state (DOS) of $\text{Bi}_2\text{Se}_3$, and another is obtained from the numerical results of energy bands difference in the diagonalized bases of total Hamiltonian $H$. The DOS calculation is done with Hartree approximation (ignoring the exchange term), as shown in Appendix.\ref{AA1}, and perform the momentum integral on the imaginary part of electron Green's function of $\text{Bi}_2\text{Se}_3$. From the shape of the DOS we may identify the gap symmetry and magnitude, but will not have momentum space resolution. The numerically obtained energy bands difference with momentum dependence complements this.  

From Eq.(\ref{warp2}) we find the desired surface state band structure of $\text{Bi}_2\text{Se}_3$, showing clear hexagonal Fermi surface at larger chemical potential as shown in Fig.\ref{sFSlat1}. This hexagonal structure is not apparent at the experimental relevant Fermi level. However, with the breaking of circular symmetry, the presence of Zeeman like $\sigma_z$ term\cite{Black}, and the crystal orientation of $\text{Bi}_2\text{Se}_3$ which is $45^{\circ}$ or $15^{\circ}$ different\cite{Wang2013} from that of BSCCO, we have generated $s$-wave, $p$-wave, and $p\pm ip$-wave pairing in additional to $d$-wave pairing on the $\text{Bi}_2\text{Se}_3$ surface. The crystal orientation difference, as demonstrated in Fig.\ref{sFSlat1}, is important for the existence of $s$-wave\cite{Wang2013} pairing. There would be no $s$-wave pairing if rotated by $15^{\circ}$ (or equivalent $45^{\circ}$ with $C_6$ symmetry in warping corrected $\text{Bi}_2\text{Se}_3$ surface state) due to the $d$-wave pairing from the BSCCO and the alignment of symmetry axes of $\text{Bi}_2\text{Se}_3$ in nodal directions. Same argument also applies to the lacking of $s$-wave pairing without the inclusion of warping terms. Aforementioned pairing amplitudes are denoted by $A_{s}$, $A_{p}$, $A_{p\pm ip}$, and $A_{d}$ and their values with different tunneling strengths $t$ are listed in Table.\ref{table4}.

\begin{figure}[h]
\centering
\subfigure[~DOS with SC]{
\includegraphics[width=.45\columnwidth]{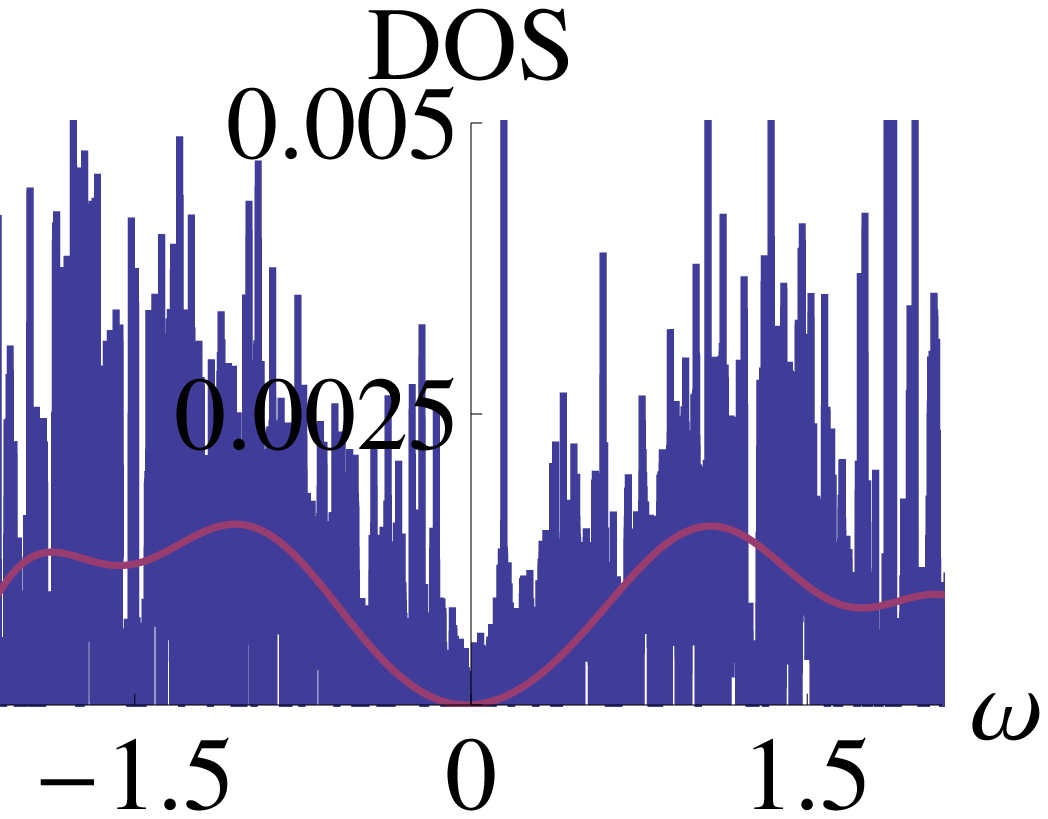}
\label{sFSDOSlat1} }
\subfigure[~SC gap]{
\includegraphics[width=.45\columnwidth]{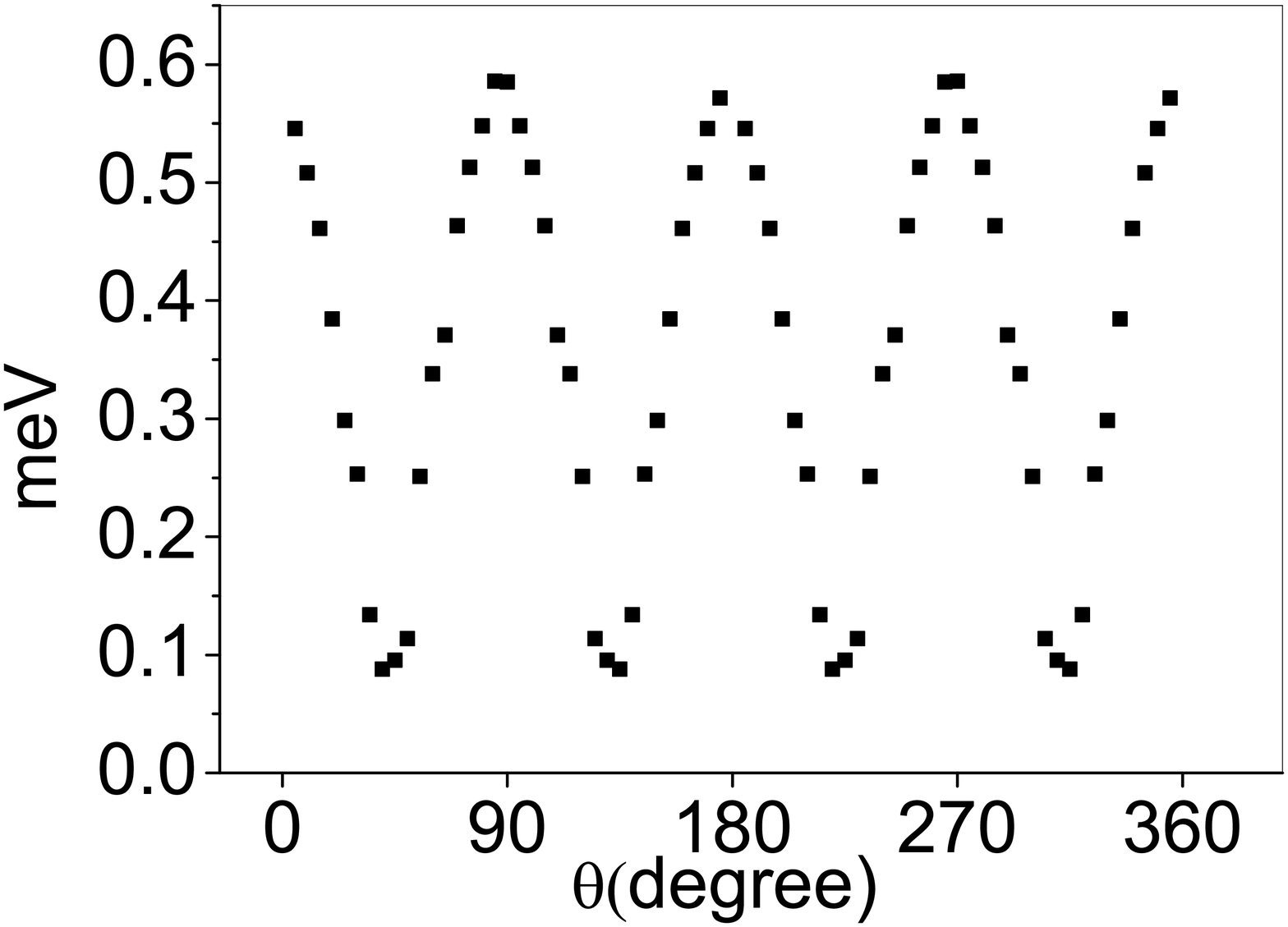}
\label{gapan3} }
\caption{(a) DOS in $\text{Bi}_2\text{Se}_3$ with $t/h=0.67$ and $|\Delta|/h\simeq0.13$. Blue line is the raw numerical data and purple line is the polynomial fit for easier identification of gap location. Frequency unit is meV. (b) Angular dependence of gap in momentum space obtained from the bandstructure in $\text{Bi}_2\text{Se}_3$/BSCCO system with warping terms. $t/h=0.67$ with angular separation of $5^{\circ}$.}
\label{sFSlatt}
\end{figure}

\begin{table}[t]
\begin{tabular}[b]{|c||c|c|c|c|c|}\hline
     $t$ &  $\bar{A}_{s}$ & $\bar{A}_{p}$ & $\bar{A}_{d}$ & $\bar{A}_{p\pm ip}$ \\ \hline
     $0.01$ &  $ -7.1698 $ & $-0.29080$ & $2.7588$ & $-2.9932i$ \\ \hline
     $0.1$ & $ -8.7897 $ & $-4.9422$ & $209.67$ & $-47.220i$ \\ \hline
       $0.67$ &  $ -364.21 $ & $-93.741$ & $8713.9$ & $-1254.4i$ \\ \hline
       $1.0$ &  $-545.43$ & $12.993$ & $17739$ & $-1381.5i$ \\ \hline
  $1.5$ &  $ -406.05 $ & $768.48$ & $32251$ & $1150.2i$ \\ \hline
  $2.0$ &  $ 1475.2 $ & $2659.3$ & $41063$ & $6947.7i$ \\ \hline
\end{tabular}
\caption{Various possible pairing amplitudes at different tunneling strengths $t$. Gap magnitude of BSCCO is fixed at $\Delta/h\simeq0.13$ with pairing amplitudes evaluated with $160\times 160$ k points. $\bar{A}_i$ is defined as $\tilde{A_i}=\bar{A}_i\times 10^{-5}$.}
\label{table4}
\end{table}

In Fig.\ref{sFSDOSlat1} we plot the DOS of $\text{Bi}_2\text{Se}_3$ surface state at tunneling strength $t/h=0.67$ with superconducting gap $\Delta/h\simeq0.13$ in BSCCO. The proximity induced gap is mainly $d$-wave-like as shown in Table.\ref{table4}, with $d$-wave gap magnitude around $0.0016h$ or $0.5$ meV (about half from DOS plot due to the $d$ wave nature). This magnitude is slightly larger\footnote{The $d$ wave gap reads out from the DOS might be larger than its actual value as the normal state bands also mix with the spectral weight of quasiparticles.} than the gap obtained from the difference of particle hole bands of $\text{Bi}_2\text{Se}_3$ nearby the Fermi level shown in Fig.\ref{gapan3}, which is around $0.3$ meV evaluated with $160\times 160$ k-points. The $s$-wave-like component read out from Fig.\ref{gapan3} is about $0.08$ meV, which gives roughly consistent ratio as $(\tilde{A_s}+\tilde{A_{p\pm ip}})/\tilde{A_d}$ shown in Table.\ref{table4}(the $p\pm ip$ also gives $s$-wave-like gap).
  
In the ARPES experiment of Ref.\onlinecite{Wang2013} seven quintuple layers (QL) of $\text{Bi}_2\text{Se}_3$ are used, and the measurement is done at the top plane which is not in contact with BSCCO. The model calculations done above is just for a single (surface) layer of $\text{Bi}_2\text{Se}_3$ in direct contact with BSCCO, assuming the thickness of $\text{Bi}_2\text{Se}_3$ is sufficient large that the Dirac modes on the two opposite surfaces are present. As we have assumed a very smooth contact between the two materials, the gap magnitude obtained, via DOS or rediagonalized band structures, in our calculations (corresponding to the proximity induced gap at the surface in contact with BSCCO) should be viewed as the maximum value. Our result shows that the gap is small ($<1$ meV at $t/h=0.67$) and mainly $d$-wave like even with the inclusion of warping and correct crystal orientation, which tends to support the absence of the proximity effect\cite{Yilmaz2014,Xu2014} in the surface states of the $\text{Bi}_2\text{Se}_3$/BSCCO system. Below we discuss other factors not included in the aforementioned discussions, and see how they modify our results. 

\section{Other relevant factors} \label{secwarpe}
We have assumed smooth interface and computed only the proximity induced gap at the surface of $\text{Bi}_2\text{Se}_3$ in direct contact with BSCCO. In the actual experiments\cite{Wang2013,Yilmaz2014,Xu2014} the ARPES were done on the opposite surface of a few quintuple layers thick $\text{Bi}_2\text{Se}_3$. From the ARPES data the Fermi level of a few QL thick $\text{Bi}_2\text{Se}_3$ actually crosses the bulk band and the bulk is not insulating. Furthermore, in the growth process, there are always some impurities generated at the bulk or surface of $\text{Bi}_2\text{Se}_3$. Last but not least, the lattice size of BSCCO is $3.8\AA$ and that of $\text{Bi}_2\text{Se}_3$ is $4.138\AA$. The latter is taken to be the same as that of BSCCO in our mean field calculations. We discuss how these factors influence our results below. 

The simplest way to consider the proximity effect on the other surface of $\text{Bi}_2\text{Se}_3$ is to assume the two surfaces are tunnel-coupled. In this case the model becomes trilayer rather than bilayer one. For moderate tunneling strength the multilayer structure always has smaller pairing amplitude compared with bilayer system (as shown in the generic model discussion in Appendix.\ref{AC}). Thus the simplest way to include this finite thickness effect of $\text{Bi}_2\text{Se}_3$ will not be able to account for the $15$ meV measured in Ref.\onlinecite{Wang2013}.

We may include the bulk band by treating it as a two bands metal on the metallic surface in contact with the BSCCO. This is because the Fermi surface of the $\text{Bi}_2\text{Se}_3$ also cut through the bulk band, and in some literature\cite{Kim2014} it is called a topological metal. In principle the bulk bands share the tunneled Cooper pairs from the superconductor with the surface band, therefore at the same tunneling strength the pairing gap at the surface decreases compared with that without the inclusion of bulk band in a self consistent calculation\footnote{The inclusion of bulk band decreases the pairing gap at the contact surface, but it helps enlarge the effective tunneling term (from insulating to metallic) between the contact surface and the surface measured in ARPES. Large interfacial tunneling term helps maintain the magnitude of the pairing gap measured in ARPES as shown in model discussion in Appendix.\ref{AC}. Here the focus is put on whether it helps enlarge the proximity induced gap at the contact surface.}. If the two bands are coupled via repulsive interactions, the pairing gaps at different bands could be different as suggested to be measured experimentally in Ref.\onlinecite{Wang2013}, where different frequencies are used to distinguish the measurement of bulk from that of surface. As both bands share the superconducting gap from BSCCO, the inclusion of bulk gap tends to lower the proximity induced gap of the surface band, rendering it even smaller than $0.6$ meV at $t/h=0.67$.

The impurities in the bulk and surfaces of $\text{Bi}_2\text{Se}_3$ could in general suppress the nodal pairings, and make $s$-wave pairing dominant for sufficient thick sample\cite{Yao2015}. The drawback is the scattering from the impurities also tend to diminish the pairing gap, making it unlikely to achieve the large ($\ge 10$ meV) proximity induced $s$-wave-like gap. Another way to enhance this $s$-wave component is to introduce an isotropic, attractive interaction as in Ref.\onlinecite{Yao2015}. However, in our zero temperature calculations, including this attractive interaction will make the $\text{Bi}_2\text{Se}_3$ superconducting without coupling to BSCCO. The undoped $\text{Bi}_2\text{Se}_3$ has not been seen to be superconducting in the experiment\footnote{$\text{Bi}_2\text{Se}_3$ is not superconducting at ambient pressure, but the high pressure leading to structural change does lead to superconductivity. See P. P. Kong et. al., J. Phys.: Condens. Matter 25, 362204 (2013). Also in $\text{Bi}_2\text{Se}_3$ made by pulsed laser deposition the $\text{Bi}_2\text{Se}_3$ becomes superconducting due to proximity effect from the superconducting $\text{Bi}$ islands grown on its surface. See P. H. Le et. al., APL Materials 2, 096105 (2014).}, and therefore we do not include this factor in our discussion. It is possible that interfacial $s$ wave superconductivity could be formed at the interface of BSCCO and $\text{Bi}_2\text{Se}_3$, and we verify this interfacial attractive interaction does enhance the $s$-wave pairing amplitude in $\text{Bi}_2\text{Se}_3$ layer but the enhancement is small (around 25$\%$) with moderate interaction strength (twice the attractive interaction strength of BSCCO at $t/h=1.3$). 

\begin{figure}
\subfigure[~DOS]{
\includegraphics[width=.45\columnwidth]{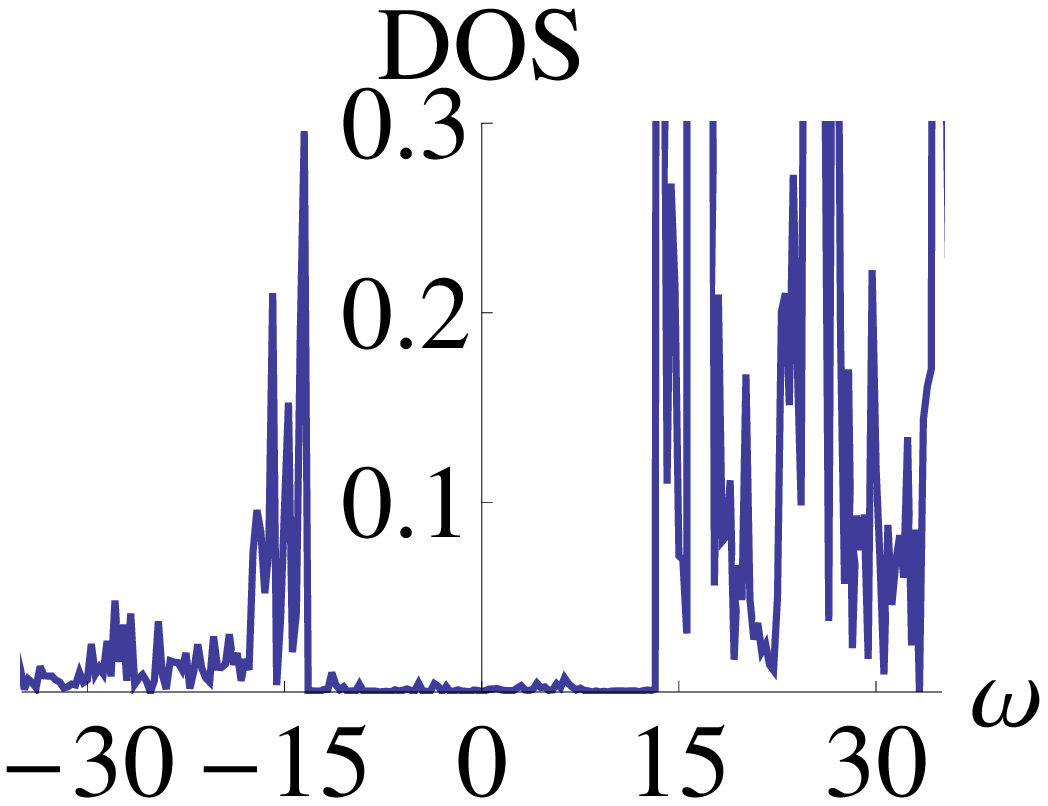}
\label{gapan5} }
\subfigure[~Comparison]{
\includegraphics[width=.45\columnwidth]{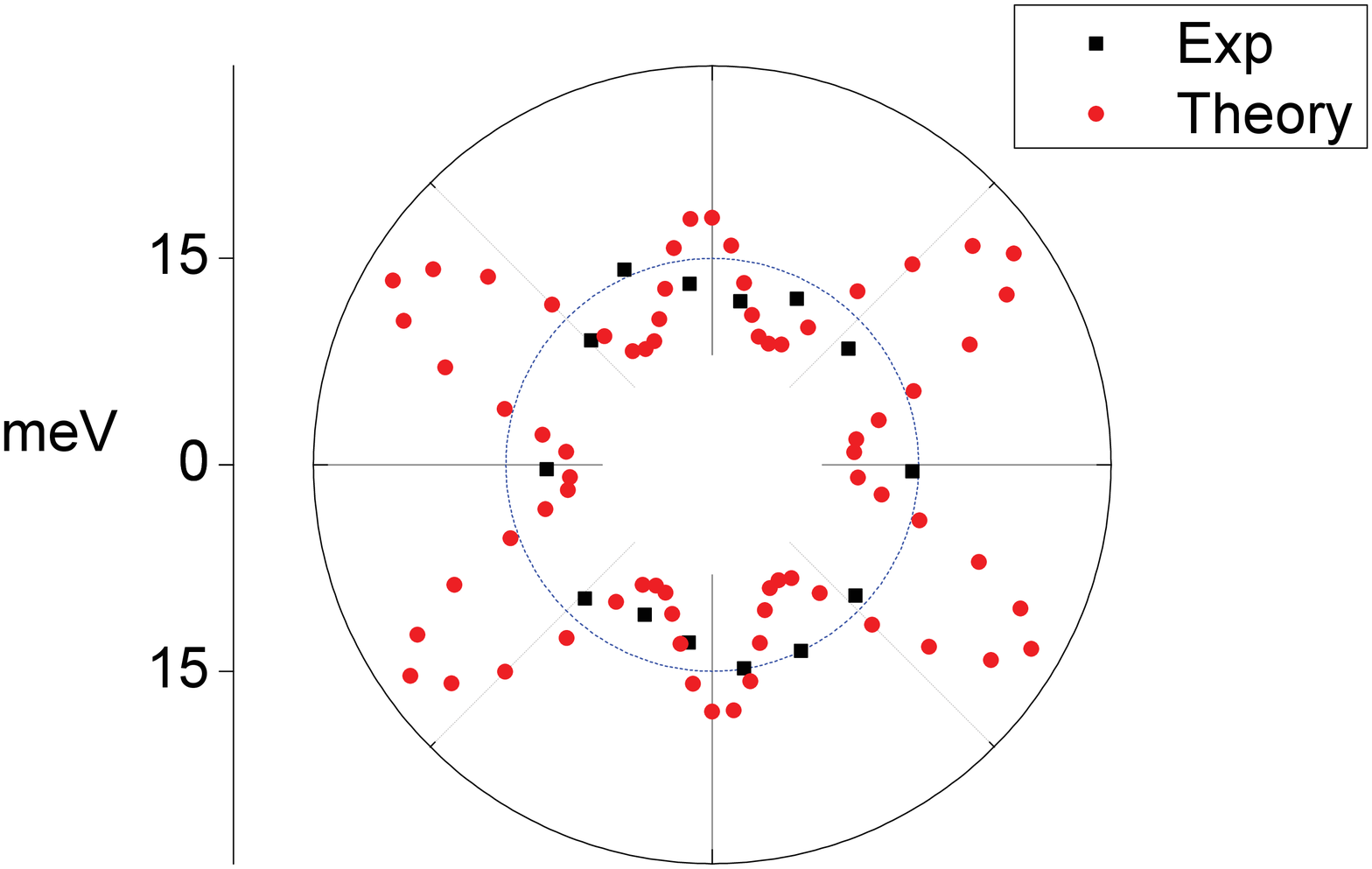}
\label{gapan6} }
\caption{(a) The DOS of $\text{Bi}_2\text{Se}_3$ at $t/h=2$ with superconducting gap in BSCCO $\Delta/h\simeq 0.13$. All other parameters are the same as those in the main text. Frequency in unit of meV. (b) Gap structure (red dots) at $t/h=2$ compared with the experimental data (black square) shown in Ref.\onlinecite{Wang2013}. }
\label{gapSOCwarp2}
\end{figure}

There are two other factors which could possibly enhance the pairing amplitudes in this system. One is the lattice mismatch issue, and the other is the possibility of forming enlarged Fermi surface in $\text{Bi}_2\text{Se}_3$ due to its growth on BSCCO (the gating issue). For different Fermi surface structure the lattice mismatch could usually lead to more overlaps in bands when taking the lattice mismatch into considerations (see the generic model discussion in Appendix.\ref{AD}). However, given the small lattice mismatch (lattice size ratio $a_2/a_1\simeq 4.1/3.8\simeq 1.08$) the major enhancement of proximity induced gap should come from the enlargement of Fermi surface on the $\text{Bi}_2\text{Se}_3$ surface. Based on our calculations, even raising to twice of the original chemical potential of $\text{Bi}_2\text{Se}_3$ does not significantly enhance the proximity induced gap magnitude (around twice and also mainly $d$-wave-like). Judging from the discrepancy of shape and gap magnitude with ARPES data shown in Ref.\onlinecite{Wang2013}, and the fact that the estimates we have here should be the theoretical upper bound for a clean interface, we tend to support the experimental results in Refs.\onlinecite{Yilmaz2014,Xu2014}. 

\section{Possible s-wave gap}\label{s wave} 
How do we explain the experimental findings in Ref.\onlinecite{Wang2013}? It is suggested\cite{Yilmaz2014} that the grounding issues could lead to incorrect readings of superconducting gaps, but this factor might not reflect the temperature dependence of the gap in a way correlated with the BSCCO observed in Ref.\onlinecite{Wang2013}. We suspect the observed $s$-wave-like gap is the result of strong hybridizations between the bands of $\text{Bi}_2\text{Se}_3$ and BSCCO due to the large tunneling amplitude. In this case the observed gap is not purely due to proximity induced superconductivity, but also the mass gap generated by single particle tunneling. 

We find for $t/h>2$ the single particle mass gap along all directions (both nodal and antinodal ones) is generated even if we turned off the superconductivity in BSCCO, and we use this as the upper bound\footnote{This critical value depends on the tight binding model of BSCCO we use, but the physics of hybrid gap, or superconducting gap mixing with single particle gap, mentioned in the main text exists for all different models. For Ref.\onlinecite{Norman1995} it is $t/h=2$ and for the tight binding model used in Ref.\onlinecite{Annett2002} it is $t/h=1.3$. For tight binding model used in Ref.\onlinecite{Hoogenboom,Fink2003} the critical $t/h$ is around $3$. This value mainly depends on the energy dispersion of BSCCO at $\Gamma$ point in momentum space.} of tunneling amplitude as no gap opening along antinodal lines is found at $T>T_c$ in Ref.\onlinecite{Wang2013}. From Table.\ref{table4} we find at $t/h=2$ the proximity induced pairing amplitudes are still dominated by $d$-wave pairing, but the shape of the gap as shown in Fig.\ref{gapan6} is similar to $s$-wave. The gap magnitude read out from the DOS of $\text{Bi}_2\text{Se}_3$ in Fig.\ref{gapan5} is around $13.4$ meV at $t/h=2$, and no gap along the antinodal direction of BSCCO is seen when the superconductivity is turned off. The lobes of this mass gap is pointing along the nodal directions of BSCCO, making it look like $s$-wave gap structure, as shown in Fig.\ref{gapan6}, when the superconductivity is turned on.

The aforementioned $s$-wave-like gap opening is consistent with the temperature dependence observed in Ref.\onlinecite{Wang2013}, i.e. no gap at $T>T_c$ and $s$-wave-like gap at $T<T_c$. At $t/h=2$ the onset of proximity induced superconductivity is estimated, from the change of DOS, to be around $60$K which is roughly consistent with the data\cite{Wang2013}. Factoring in the aforementioned lattice mismatch, enlarged Fermi surfaces, possible $s$-wave pairing interaction at the interface, could possibly give rise to the scale and features similar to the experimental results in Ref.\onlinecite{Wang2013}. We emphasize that the $s$-wave-like structure shown in Fig.\ref{gapSOCwarp2} is the combined results of bands hybridization between $\text{Bi}_2\text{Se}_3$ and BSCCO and the $d$-wave superconductivity from BSCCO. The proximity induced pairing amplitude in $\text{Bi}_2\text{Se}_3$ is still dominated by $d$-wave. This scenario is different from the dominant $s$-wave pairing discussed in Ref.\onlinecite{Yao2015}, in which the $s$-wave pairing gap is completely from the proximity induced superconductivity.

\section{Conclusion}\label{discon}
 In this paper, we use the mean-field approach to compute the superconducting
pairing amplitudes and gap magnitude in a superconducting-metal interface. We find the pairing amplitude is mainly $d$-wave except at small $t$ ($t/h \le 0.01$ with $h$ the nearest neighbor hopping amplitude in BSCCO). With reasonably large $t$ ($2> t/h \ge 0.1$), the gap structure, obtained from our DOS and band calculations, is mainly $d$-wave with gap magnitude less than $1$ meV.
 
 We suggest one possible reason why the large $s$ wave like gap ($\sim 15$ meV) is observed in Ref.\onlinecite{Wang2013} while the other two similar setups\cite{Yilmaz2014,Xu2014} show no significant ($>5$ meV) proximity induced gap. We argue the gap in Ref.\onlinecite{Wang2013} is not purely due to superconductivity, but rather contains a large component due to a large tunneling amplitude($t/h\ge 2$). The pairing amplitude evaluated at large $t$ are still dominated by the $d$-wave channel, and its mixing with the single-particle gap makes it more $s$-wave-like, as shown in Fig.\ref{gapan6}. Thus, we propose that sample-to-sample variations in interfacial coupling between $\text{Bi}_2\text{Se}_3$ and BSCCO explain the discrepancy between the observed gaps, and also that the induced superconductivity is much smaller than would be naively concluded from the observations of Ref.\onlinecite{Wang2013}.

 A piece of supporting evidence for this claim, as also pointed out in Ref.\onlinecite{Yilmaz2014} and \onlinecite{Xu2014}, is the lack of superconducting coherence peak in the observed ARPES spectrum in Ref.~\onlinecite{Wang2013}. Furthermore, it is claimed\cite{Wang2013} that the bulk conducting bands of $\text{Bi}_2\text{Se}_3$ are also superconducting, but with a much smaller gap. As the bulk bands are further away from the quasiparticle bands in BSCCO and their effective tunneling amplitudes are smaller due to smaller wave function overlaps, it is possible that the gap induced in the bulk bands is dominated by the $d$-wave superconducting gap and the measured gap might be mainly from proximity induced superconductivity. More detailed measurements of the bulk spectrum should shed light on this issue. This case also serves as a cautionary example: measuring the temperature dependence of a gap at an interface is not sufficient to show that the gap is superconducting in origin.

\acknowledgments
The authors acknowledge useful discussions with Peng-Jen Chen (Academia Sinica, Taiwan) and thank Chuck-Hou Yee (Rutgers University, U.S.A.) for improving our manuscript. WJL and TKL acknowledge the support by Taiwan's MOST (No.103-2119-M001-011-MY2). SPC thanks the support by Taiwan's MOST (No.103-2112-M-001-035-MY3) and the Simons Foundation for his stay at Aspen, C.O., U.S.A., where part of this work is done.\\
   
\appendix
\section{Perturbative results for proximity induced pairing amplitude}\label{AA}
We derive the analytical, perturbative results of superconducting proximity effect for some generic bilayer and trilayer systems using Green's functions formalism. The perturbative term is the attractive interaction in the superconducting layer, and we further assume the tunneling amplitude is small to simplify our results. The form of self energy in the Green's function is assumed to take the Hartree form (neglecting exchange interactions) with mean field approximations. The real part of self energy just shifts or renormalizes the chemical potential. Under these approximations the self consistent equation of the superconducting layer is the same as the uncoupled one, and the leading term of pairing amplitude in the first metallic layer is proportional to $t^2$. The generic model Hamiltonian we considered in the Appendix is:
\bea\nn
&&H=\sum_{\vec{k},l,\sigma,\sigma'}(\epsilon_l(\vec{k})-\mu_l)_{\sigma\sigma'}c_{\vec{k}l\sigma}^{\dagger}c_{\vec{k}l\sigma'}+\sum_{\vec{k},i<j,\sigma} t_{ij}\\\nn
&&\times(c_{\vec{k}i\sigma}^{\dagger} c_{\vec{k}j\sigma}+c_{\vec{k}j\sigma}^\dagger c_{\vec{k}i\sigma})+\sum_{\vec{k},\vec{k'},\vec{q},l,l'}V_{ll'}(\vec{q})c_{\vec{k}l\uparrow}^{\dagger}c_{(\vec{k}-\vec{q})l\uparrow}\\\label{mult}
&&\times c_{(\vec{k'}-\vec{q})l'\downarrow}^{\dagger}c_{\vec{k'}l'\downarrow}
\eea
Here $l,i,j$ range from $1,2,3,\ldots,N$ with $1$ denote electron/hole operators on the superconducting layer and $2$, $3$ (say $N=3$) denote two bands in one metallic layer or two metallic layers. In this model Hamiltonian we consider the usual form of density density interactions. Below we show explicitly the calculations for three cases: Bilayer, trilayer, and metallic layer with spin orbit interactions.

\subsection{Two layers with single band}\label{AA1}
Here we perform perturbative calculations via the Green's function formalism. Using the four component Nambu spinor bases $\Psi_{\vec{k}}=\big(c_{\vec{k}1\uparrow}^{\dagger} \quad c_{-\vec{k}1\downarrow} \quad c_{\vec{k}2\uparrow}^{\dagger} \quad c_{-\vec{k}2\downarrow}\big)^T$, the Green's function is given by
\bea
G(\vec{k},\tau)=-\langle T_\tau \Psi(\vec{k},\tau)\Psi^{\dagger}(\vec{k},0)\rangle
\eea
For $V_l(\vec{q})=0$ the retarded Green's function $G_0(\vec{k},\omega)$ takes the form:
\bw
\bea\label{g0}
G_0(\vec{k},\omega)= \left( \begin{array}{cccc} \omega-\bar{\epsilon}_1(\vec{k})+i\eta &  0 & t & 0 \\  0 & \omega+\bar{\epsilon}_1(\vec{k})+i\eta  & 0 & -t \\t& 0 & \omega-\bar{\epsilon}_2(\vec{k})+i\eta  & 0 \\ 0 & -t & 0 & \omega+\bar{\epsilon}_2(\vec{k})+i\eta  \end{array}\right)^{-1}\;
\eea
\ew
with $\bar{\epsilon}_l(\vec{k})=\epsilon_l(\vec{k})-\mu_l$.
Turning on $V_l(\vec{q})\neq 0$ gives self energy correction on the bare Green's function $G_0(\vec{k},\omega)$ from perturbations in $V_l(\vec{q})$. We take $\vec{q}=\vec{k}+\vec{k'}$ as in the BCS theory and denote $V_l(\vec{q})=V_l(\vec{k},\vec{k'})=V_l$ in the following. The perturbation in $V_l$ gives:
\bea\label{gko}
G(\vec{k},\omega)^{-1}=G_0(\vec{k},\omega)^{-1}-\Sigma(\vec{k},\omega). 
\eea
Based on first order perturbation in $V_l(\vec{q})$, the Hartree corrected\cite{Kamar} $\Sigma(\vec{k},\omega)\simeq \Sigma_H $ takes the following form
\bea\nn
\Sigma_H=\left( \begin{array}{cccc} \Sigma_1(\vec{k},\omega) & S_1(\vec{k},\omega) &0 &0\\\ S_1^{\dagger}(\vec{k},\omega)& -\Sigma_1(-\vec{k},-\omega) &0 &0 \\0 &0 & \Sigma_2(\vec{k},\omega) & S_2(\vec{k},\omega) \\ 0 &0 & S_2^{\dagger}(\vec{k},\omega)& -\Sigma_2(-\vec{k},-\omega) \end{array}\right)\;
\eea
with $\rho(\vec{k},\omega)\equiv -\frac{1}{\pi}\Im(G(\vec{k},\omega))$, $f_i(\omega)=\theta(\mu_i-\omega)$ the Fermi-Dirac distribution at zero temperature, and 
\bea\label{s1}
&&\Sigma_1(\vec{k},\omega)\simeq \sum_{\vec{k'}}\frac{V_1}{2}\int_{-\infty}^{\infty}d\omega \rho_{11}(\vec{k'},\omega)f_1(\omega)\\\label{s2}
&&\Sigma_2(\vec{k},\omega)\simeq \sum_{\vec{k'}}\frac{V_2}{2}\int_{-\infty}^{\infty}d\omega \rho_{33}(\vec{k'},\omega)f_2(\omega)\\\label{s3}
&&S_1(\vec{k},\omega)\simeq \sum_{\vec{k'}}V_1 \int_{-\infty}^{\infty}d\omega \rho_{12}(\vec{k'},\omega)f_1(\omega)\\\label{s4}
&&S_2(\vec{k},\omega)\simeq \sum_{\vec{k'}}V_2 \int_{-\infty}^{\infty}d\omega \rho_{34}(\vec{k'},\omega)f_2(\omega)
\eea
Eq.(\ref{s1})-Eq.(\ref{s4}) is named Hartree approximation, and under this approximation the $\Sigma_i$ and $S_i$ terms have momentum but no frequency dependence. 
For a given set of $V_i$, the $\Sigma_i$ and $S_i$ can be obtained via iterations. We take $V_1<0$ and $V_2=0$ to simulate the case of first layer being superconductor and second layer being normal metal before placed together. Note that
\bea
\langle c_{\vec{k}1\uparrow}^{\dagger}c_{-\vec{k}1\downarrow}^{\dagger}\rangle=\int d\omega \frac{-1}{\pi}\Im(G(\vec{k},\omega)_{12})\\
\langle c_{\vec{k}2\uparrow}^{\dagger}c_{-\vec{k}2\downarrow}^{\dagger}\rangle=\int d\omega \frac{-1}{\pi}\Im(G(\vec{k},\omega)_{34})
\eea
with $G(\vec{k},\omega)_{ij}$ denote $i$, $j$ component of the $4\times 4$ Green's function. The momentum sum, along with appropriate symmetry factor $f_i$ defined in Eq.(\ref{eq6}), in above equations give pairing amplitude mentioned in the main text. The DOS of metallic layer $\rho(\omega)$ is computed by
\bea\label{DOS}
 \rho(\omega)=\frac{1}{\pi}\sum_{i=3,4}\lim_{\eta\rightarrow 0}\Im\{\int \frac{d\vec{k}}{(2\pi)^2}G^R_{ii}(\omega+i\eta,\vec{k})\}
\eea 
Using assumptions Eq.(\ref{s1})-Eq.(\ref{s4}) and the conditions $V_1\neq 0$, $V_2=0$ in Eq.(\ref{gko}) we have
\bw
\bea 
G(\vec{k},\omega)_{12}&=&\frac{S_1(\vec{k},\omega)(\tilde{\omega}^2-\bar{\epsilon}_2(\vec{k})^2)}{(\tilde{\omega}^2-\bar{\epsilon}_2(\vec{k})^2) ((\bar{\epsilon}_1(\vec{k})+\Sigma_1(\vec{k},\omega)^2+S_1(\vec{k},\omega)^2-\tilde{\omega}^2)+2t^2(\tilde{\omega}^2+\bar{\epsilon}_2(\vec{k})(\bar{\epsilon}_1(\vec{k})+\Sigma_1(\vec{k},\omega)))-t^4}\\
G(\vec{k},\omega)_{34}&=&\frac{-t^2S_1(\vec{k},\omega)}{(\tilde{\omega}^2-\bar{\epsilon}_2(\vec{k})^2) ((\bar{\epsilon}_1(\vec{k})+\Sigma_1(\vec{k},\omega))^2+S_1(\vec{k},\omega)^2-\tilde{\omega}^2)+2t^2(\tilde{\omega}^2+\bar{\epsilon}_2(\vec{k})(\bar{\epsilon}_1(\vec{k})+\Sigma_1(\vec{k},\omega)))-t^4}
\eea
\ew
with $\tilde{\omega}=\omega+i\eta$ for retarded Green's function.
For $t=0$ we have $\langle c_{\vec{k}2\uparrow}^{\dagger}c_{-\vec{k}2\downarrow}^{\dagger}\rangle=0$ and 
\bea\nn
\langle c_{\vec{k}1\uparrow}^{\dagger}c_{-\vec{k}1\downarrow}^{\dagger}\rangle&=&\frac{-S_1(\vec{k},\omega)}{2\sqrt{(\bar{\epsilon}_1(\vec{k})+\Sigma_1(\vec{k},\omega))^2+S_1(\vec{k},\omega)^2}}\\\label{selfcon}
&\equiv& \frac{-S_1(\vec{k},\omega)}{2E_1(\vec{k})}
\eea
The single particle self energy correction terms $\Sigma_i$ can be absorbed into the renormalized chemical potential $\mu_i$, and we remove this term in the following expressions for simplicity.

By combining Eq.(\ref{selfcon}) with Eq.(\ref{s3}), we get the self consistent gap equation in the BCS theory by identifying superconducting gap $\Delta_1(\vec{k})=S_1(\vec{k},\omega)$. For small but finite $t$
we take the leading $t^2$ correction in the evaluation of frequency integral for $G(\vec{k},\omega)_{34}$. Under this small $t$ approximation the pole structure is the same
as in Eq.(\ref{selfcon}), and we have:
\bea\label{selfcon2}
\langle c_{\vec{k}2\uparrow}^{\dagger}c_{-\vec{k}2\downarrow}^{\dagger}\rangle\simeq\frac{t^2 S_1(\vec{k},\omega)}{2E_1(\vec{k})\big(E_1(\vec{k})^2-\bar{\epsilon}_2(\vec{k})^2\big) }.
\eea
Note that it is straightforward to obtain full analytic results without this small $t$ (tunneling strength) approximation, but the results are less illuminating.
We stick with the Hartree and small $t$ approximations in this Appendix \ref{AA} to illustrate the main idea. 

From Eq.(\ref{selfcon2}) the rough estimate gives the second layer superconductivity gap magnitude $|\Delta_2|$:
\bea\label{prox}
|\Delta_2|\simeq k_BT_{c_2}\le \frac{t^2}{|\epsilon_2^2-E_1^2|}|\Delta_1|\simeq \frac{t^2 k_B T_{c_1}}{|\epsilon_2^2-E_1^2|}
\eea
 Here we use $|\Delta_2|\le |\sum_{\vec{k}}V_1\langle c_{\vec{k}2\uparrow}^{\dagger}c_{-\vec{k}2\downarrow}^{\dagger}\rangle|$, and $E_i$ and $\epsilon_i$ as the average of $E_i(\vec{k})$ and $\bar{\epsilon}_i(\vec{k})$ in momentum space as a rough estimate. From this estimate, we know for $\bar{\epsilon}_2(\vec{k})$ very different from $\bar{\epsilon}_1(\vec{k})$ ("mismatched" Fermi surface) the denominator in Eq.(\ref{prox}) increases, giving rise to smaller gap magnitude as expected. The gap symmetry of the second layer is determined by both the first layer gap symmetry and the Fermi surfaces from both layers. We do not use this estimate in the main text, but Eq.(\ref{prox}) gives naive intuitions why the mismatched Fermi surfaces give smaller proximity induced pairing amplitude.

\subsection{Two bands or three layers}\label{AA3}
For two metallic bands or two metallic layers we follow the same definition of generalized Green's function and extend the $4\times 4$ bases to $6\times 6$ ones to accommodate this extra degree of freedom. For $V_1(\vec{q})=0$ we have the retarded Green's function $G_0(\vec{k},\omega)$ taking the form:
\bw
\bea\label{g01}
G_0(\vec{k},\omega)= \left( \begin{array}{cccccc} \omega-\bar{\epsilon}_1(\vec{k})+i\eta &  0 & t_{12} & 0 & t_{13} & 0\\  0 & \omega+\bar{\epsilon}_1(\vec{k})+i\eta  & 0 & -t_{12} & 0 & -t_{13} \\t_{12} & 0 & \omega-\bar{\epsilon}_2(\vec{k})+i\eta  & 0 & t_{23} & 0 \\ 0 & -t_{12} & 0 & \omega+\bar{\epsilon}_2(\vec{k})+i\eta  & 0 & -t_{23} \\ t_{13} & 0 & t_{23} & 0 & \omega-\bar{\epsilon}_3(\vec{k})+i\eta & 0\\ 0 & -t_{13} & 0 & -t_{23} & 0 & \omega+\bar{\epsilon}_3(\vec{k})+i\eta \end{array}\right)^{-1}\;
\eea
\ew
For $\Sigma(\vec{k},\omega)\simeq \Sigma_H$ as in previous case we have 
\bw
\bea\nn
&&G(\vec{k},\omega)_{12}=S_1(\vec{k},\omega)\Big(t_{23}^4-2t_{23}^2(\tilde{\omega}+\bar{\epsilon}_2(\vec{k})\bar{\epsilon}_3(\vec{k}))^2
+(\tilde{\omega}-\bar{\epsilon}_2(\vec{k}))^2 (\tilde{\omega}-\bar{\epsilon}_3(\vec{k}))^2\Big)/De(\vec{k},\omega)\\\nn
&&G(\vec{k},\omega)_{34}=S_1(\vec{k},\omega)\left((t_{13}t_{23}-t_{12}(\tilde{\omega}-\bar{\epsilon}_3(\vec{k}))(t_{13}t_{23}+t_{12}(\tilde{\omega}+\bar{\epsilon}_3(\vec{k}))\right)/De(\vec{k},\omega)\\\nn
&&G(\vec{k},\omega)_{56}=S_1(\vec{k},\omega)\left((t_{12}t_{23}-t_{13}(\tilde{\omega}-\bar{\epsilon}_2(\vec{k}))(t_{12}t_{23}+t_{13}(\tilde{\omega}+\bar{\epsilon}_2(\vec{k}))\right)/De(\vec{k},\omega)\\\nn
&&De(\vec{k},\omega)\equiv t_{13}^4(\bar{\epsilon}_2(\vec{k})^2-\tilde{\omega}^2)+4t_{12}^3t_{13}t_{23}\bar{\epsilon}_3(\vec{k})+t_{12}^4(\bar{\epsilon}_3(\vec{k})^2-\tilde{\omega}^2)+2t_{13}^2\Big(t_{23}^2\big(-\tilde{\omega}^2+\bar{\epsilon}_2(\vec{k})(\bar{\epsilon}_1(\vec{k})+\Sigma_1(\vec{k},\omega))\big)\\\nn
&&+(\tilde{\omega}^2-\bar{\epsilon}_2(\vec{k})^2)\big(\tilde{\omega}^2+\bar{\epsilon}_3(\vec{k})(\bar{\epsilon}_1(\vec{k})+\Sigma_1(\vec{k},\omega))\Big)-4t_{12}t_{13}t_{23}\big((\tilde{\omega}^2-t_{13}^2)\bar{\epsilon}_2(\vec{k})+\tilde{\omega}^2\bar{\epsilon}_3(\vec{k})+(\bar{\epsilon}_1(\vec{k})+\Sigma_1(\vec{k},\omega))\\\nn
&&\times(\tilde{\omega}^2-t_{23}^2+\bar{\epsilon}_2(\vec{k})\bar{\epsilon}_3(\vec{k}))\big)+2t_{12}^2\big(t_{23}^2(-\tilde{\omega}^2+(\bar{\epsilon}_1(\vec{k})+\Sigma_1(\vec{k},\omega))\bar{\epsilon}_3(\vec{k}))+t_{13}^2(-\tilde{\omega}^2+2t_{23}^2+\bar{\epsilon}_2(\vec{k})\bar{\epsilon}_3(\vec{k}))+(\tilde{\omega}^2\\\nn
&&+(\bar{\epsilon}_1(\vec{k})+\Sigma_1(\vec{k},\omega))\bar{\epsilon}_2(\vec{k}))(\tilde{\omega}^2-\bar{\epsilon}_3(\vec{k})^2)\big)+\big((S_1(\vec{k},\omega)^2-\tilde{\omega}^2)+(\bar{\epsilon}_1(\vec{k})+\Sigma_1(\vec{k},\omega))^2)(t_{23}^4-2t_{23}^2(\tilde{\omega}^2\\\nn
&&+\bar{\epsilon}_2(\vec{k})\bar{\epsilon}_3(\vec{k}))+(\tilde{\omega}^2-\bar{\epsilon}_2(\vec{k})^2)(\tilde{\omega}^2-\bar{\epsilon}_3(\vec{k})^2))\big)
\eea
\ew
The tunneling strength $t_{ij}$ depends on the overlap integrals, i.e. the symmetry of the respective eigenstates. For the special case of $t_{13}=0$ (such as the trilayer case, where there is no direct tunneling between the first superconductor layer and third metallic layer) but finite
and small $t_{12}$ and $t_{23}$, we take the leading order in $t_{ij}$ for $G(\vec{k},\omega)$. With this approximation, we carry out the frequency integral and obtain: 
\bea
&&\langle c_{\vec{k}_1\uparrow}^{\dagger}c_{-\vec{k}_1\downarrow}^{\dagger}\rangle\simeq \frac{-S_1(\vec{k},\omega)}{2E_1(\vec{k})}\\\label{a23}
&&\langle c_{\vec{k}_2\uparrow}^{\dagger}c_{-\vec{k}_2\downarrow}^{\dagger}\rangle\simeq \frac{t_{12}^2 S_1(\vec{k},\omega)}{2E_1(\vec{k})\big(E_1(\vec{k})^2-\bar{\epsilon}_2(\vec{k})^2\big) }\\\nn
&&\langle c_{\vec{k}_3\uparrow}^{\dagger}c_{-\vec{k}_3\downarrow}^{\dagger}\rangle \\\label{a24}&&\simeq \frac{-S_1(\vec{k},\omega)t_{12}^2t_{23}^2}{2E_1(\vec{k})\big(E_1(\vec{k})^2-\bar{\epsilon}_2(\vec{k})^2\big)\big(E_1(\vec{k})^2-\bar{\epsilon}_3(\vec{k})^2\big)}
\eea

Similar to the bilayer case, we obtain the rough estimate for proximity induced gap magnitude on the third layer: $|\Delta_3|\le |\sum_{\vec{k}}V_1\langle c_{\vec{k}3\uparrow}^{\dagger}c_{-\vec{k}3\downarrow}^{\dagger}\rangle|\simeq \frac{t_{23}^2}{|E_1^2-\epsilon_3^2|}|\Delta_2|\simeq \frac{t_{12}^2t_{23}^2}{|(E_1^2-\epsilon_2^2)(E_1^2-\epsilon_3^2)|}|\Delta_1|$. 

The same formulation can be extended to the case of two bands in a single metallic layer. In this case, we perform the same calculations but with $t_{23}=0$ and nonzero but small $t_{12}$ and $t_{13}$. The label $2$ and $3$ here denote the two bands in the metallic layer. We will not present the calculations here, but just mention that it is a straightforward extension of current formula.

\subsection{Metallic layer with spin orbit interactions}\label{AA2}
For two dimensional electron gas in a symmetric quantum well, such as the heterostructure in the bilayer system discussed above, or the surface state of three dimensional topological insulator, the anisotropy or asymmetry in general leads to some type of spin orbit interactions. To include this factor, we consider a generic linear momentum dependent spin orbit coupling in our two dimensional metallic layer:
\bea\nn
H_{so}&\simeq& \alpha k_x\sigma_y+\beta k_y\sigma_x=\frac{1}{2}(\alpha+\beta)(k_x\sigma_y+k_y\sigma_x)\\\label{spob}
&+&\frac{1}{2}(\alpha-\beta)(k_x\sigma_y-k_y\sigma_x)
\eea

The term times $(\alpha+\beta)$ is named Dresslhaus effect, often occurring in systems lacking reflection symmetry. The other term associated with $(\alpha-\beta)$ is called Bychkov-Rashba effect which happens when inversion symmetry is broken. To explicitly incorporate the spin degree of freedom we use four component Nambu bases on a single layer: $\Psi_{\vec{k},l}=\big(c_{\vec{k}l\uparrow}\quad c_{\vec{k}l\downarrow}\quad c_{-\vec{k}l\downarrow}^{\ast}\quad -c_{-\vec{k}l\uparrow}^{\ast}\big)^T$. Here $l=1,2$ denotes layer index and this basis is chosen as the basis for direct product of spin and electron hole space. For real $\Delta_1$, the matrix form of the Hamiltonian describing the superconducting layer ($H_S$), normal metal layer ($H_N$), and tunneling term between the two $H_t$, in this four component basis are:
\bea\nn
&&H_S=\left(\epsilon_1(\vec{k})-\mu_1\right)\tau_z\otimes\sigma_0+\Delta_1\tau_x\otimes\sigma_0 ,\\\nn
&&H_N=\left(\epsilon_2^{(0)}(\vec{k})-\mu_2\right)\tau_z\otimes\sigma_0+\tau_z\otimes\left(\alpha k_{x}\sigma_y+\beta k_y\sigma_x\right),\\\label{somat}
&&H_t=t\tau_z\otimes\sigma_0.
\eea

Note that we use $\epsilon_2^{(0)}(\vec{k})$ to denote the diagonal (in spin space) part of the $H_N$, and $\epsilon_2(\vec{k})-\mu_2=\epsilon_2^{(0)}(\vec{k})\pm \sqrt{|\alpha k_x|^2+|\beta k_y|^2}-\mu_2$ is used to denote the eigenvalue of $H_N$ as in other sections. Furthermore, in Eq.(\ref{somat}) we have assumed $\Delta_1=\Delta(\vec{k})=\Delta(-\vec{k})$ by placing the conventional even pairing superconductor ($s$ or $d$ wave superconductor) on top of the metallic layer. The $4\times 4$ matrix forms of Eq.(\ref{somat}) are:
\[ H_S=\left(
  \begin{array}{cccc}
    \bar{\epsilon}_1(\vec{k}) & 0 & \Delta_1(\vec{k}) & 0 \\
    0 & \bar{\epsilon}_1(\vec{k}) & 0 & \Delta_1(-\vec{k}) \\
    \Delta_1^{\ast}(\vec{k}) & 0 & -\bar{\epsilon}_1(\vec{k}) & 0 \\
    0 & \Delta_1^{\ast}(-\vec{k}) & 0 & -\bar{\epsilon}_1(\vec{k}) \\
  \end{array}
\right)\]\[
 H_N=\left(
  \begin{array}{cccc}
    \bar{\epsilon}_2(\vec{k}) & \epsilon_{so}(\vec{k}) & 0 & 0 \\
    \epsilon_{so}(\vec{k})^{\ast} & \bar{\epsilon}_2(\vec{k}) & 0 & 0 \\
    0 & 0 & -\bar{\epsilon}_2(\vec{k}) & -\epsilon_{so}(\vec{k}) \\
    0 & 0 & -\epsilon_{so}(\vec{k})^{\ast} & -\bar{\epsilon}_2(\vec{k}) \\
  \end{array}
\right)
\]
Here $\epsilon_{so}(\vec{k})\equiv\beta k_y+i\alpha k_x$. The tunneling term $H_t$ with constant tunneling amplitude $t$ takes the form:
\[ H_t=\left(
  \begin{array}{cccc}
    t & 0 & 0 & 0 \\
    0 & t & 0 & 0 \\
    0 & 0 & -t & 0 \\
    0 & 0 & 0 & -t \\
  \end{array}
\right)
\]
 This $4\times 4$ matrix for $H_t$ connects the bases of $\Psi_{\vec{k},1}$ and $\Psi_{\vec{k},2}$, with the assumption of momentum conservation. In this paper $\Delta_1(-\vec{k})=\Delta_1(\vec{k})$ as our superconductor is either $s$ wave or $d$ wave type. The full Hamiltonian, given by 
\[ H=\left(
  \begin{array}{cc}
     H_S & H_t \\
    H_t^{\ast} & H_N  \\
  \end{array}
\right),
\]
 is described by a $8\times 8$ matrix or in the
$(\Psi_{\vec{k},1},\Psi_{\vec{k},2})$ bases. Assuming $t$ is real and denoting $\tilde{\omega}=\omega+i\eta$ as before, the matrix element of the retarded Green's function $G=(\omega-H+i\eta)^{-1}$ related to second layer pairing amplitude $A_2$ are 
\bw
\bea 
&&G(\vec{k},\omega)_{57}=-G(\vec{k},\omega)_{68}\\\nn
&&=\Delta_1(\vec{k})t^2\left(t^4-2t^2\left(\tilde{\omega}^2+\bar{\epsilon}_1(\vec{k})\bar{\epsilon}_2(\vec{k})\right)+\left(|\Delta_1(\vec{k})|^2-\tilde{\omega}^2+\bar{\epsilon}_1(\vec{k})^2\right)\left(|\alpha k_x|^2+|\beta k_y|^2-\tilde{\omega}^2+\bar{\epsilon}_2(\vec{k})^2\right)\right)/De_1(\vec{k},\omega),\\
&&G(\vec{k},\omega)_{58}=2\Delta(\vec{k})t^2(i\alpha k_x+\beta k_y)\left(t^2\bar{\epsilon}_1(\vec{k})-\left(|\Delta_1(\vec{k})|^2-\tilde{\omega}^2+\bar{\epsilon}_1(\vec{k})^2\right) \bar{\epsilon}_2(\vec{k})\right)/De_1(\vec{k},\omega),\\\nn
&&De_1(\vec{k},\omega)=\big((t^4-2t^2\tilde{\omega}^2+(\Delta(\vec{k})^2-\tilde{\omega}^2)(|\alpha k_x|^2+|\beta k_y|^2-\tilde{\omega}^2)\big)^2+4t^2\epsilon_1(\vec{k}) \epsilon_2(\vec{k})(\bar{\epsilon}_1(\vec{k})^2+\Delta(\vec{k})^2-\tilde{\omega}^2)\\\nn
&&\times (|\alpha k_x|^2+|\beta k_y|^2+\tilde{\omega}^2-\bar{\epsilon}_2(\vec{k})^2)-4t^2\bar{\epsilon}_1(\vec{k})\bar{\epsilon}_2(\vec{k})(t^4-2t^2\tilde{\omega}^2)+(\Delta(\vec{k})^2-\tilde{\omega}^2)\bar{\epsilon}_2(\vec{k})^2(2(t^4-2t^2\tilde{\omega}^2-(\Delta(\vec{k})^2-\tilde{\omega}^2)\\\nn
&&\times(|\alpha k_x|^2+|\beta k_y|^2+\tilde{\omega}^2-\bar{\epsilon}_2(\vec{k})^2)))+\bar{\epsilon}_1(\vec{k})^4(|\alpha k_x|^2+|\beta k_y|^2+\tilde{\omega}^2-\bar{\epsilon}_2(\vec{k})^2)^2+2\bar{\epsilon}_1(\vec{k})^2\big((|\alpha k_x|^2+|\beta k_y|^2)(-t^4\\\nn
&&+\Delta(\vec{k})^2(|\alpha k_x|^2+|\beta k_y|^2))-(2\Delta(\vec{k})^2(|\alpha k_x|^2+|\beta k_y|^2)+(t^2+|\alpha k_x|^2+|\beta k_y|^2)^2)\tilde{\omega}^2+(\Delta(\vec{k})^2+2(|\alpha k_x|^2+|\beta k_y|^2+t^2))\\\nn
&&\times\tilde{\omega}^4-\tilde{\omega}^6+(3t^4-2t^2\tilde{\omega}^2-2(\Delta(\vec{k})^2-\tilde{\omega}^2)(|\alpha k_x|^2+|\beta k_y|^2+\tilde{\omega}^2))\bar{\epsilon}_2(\vec{k})^2+(\Delta(\vec{k})^2-\tilde{\omega}^2)\bar{\epsilon}_2(\vec{k})^4\big).
\eea
\ew
This $G(\vec{k},\omega)_{57}=-G(\vec{k},\omega)_{68}$ has to do with the fact that there is no $\sigma_z$ term in this spin orbit metallic Hamiltonian (which is not the case in the main text). Following the same Hartree and small $t$ approximations as in Appendix \ref{AA1} we get
\bw
\bea\label{a17}
&&\langle c_{\vec{k}2\uparrow}^{\dagger}c_{-\vec{k}2\downarrow}^{\dagger}\rangle\simeq\frac{-\Delta_1(\vec{k})t^2(|\alpha k_x|^2+|\beta k_y|^2+\bar{\epsilon}_2(\vec{k})^2-E_1(\vec{k})^2)}{2E_1(\vec{k})\left((E_1(\vec{k})^2-|\alpha k_x|^2-|\beta k_y|^2)^2-2(E_1(\vec{k})^2+|\alpha k_x|^2+|\beta k_y|^2)\bar{\epsilon}_2(\vec{k})^2+\bar{\epsilon}_2(\vec{k})^4\right)}\\\label{a18}
&&\langle c_{\vec{k}2\uparrow}^{\dagger}c_{-\vec{k}2\uparrow}^{\dagger}\rangle\simeq\frac{\Delta_1(\vec{k})t^2(i\alpha k_x+\beta k_y)\bar{\epsilon}_2(\vec{k})}{E_1(\vec{k})\left((E_1(\vec{k})^2-|\alpha k_x|^2-|\beta k_y|^2)^2-2(E_1(\vec{k})^2+|\alpha k_x|^2+|\beta k_y|^2)\bar{\epsilon}_2(\vec{k})^2+\bar{\epsilon}_2(\vec{k})^4\right)}
\eea
\ew

\begin{table}[t]
    \begin{tabular}[b]{|c||c|c|c|}\hline
      $(\alpha,\beta)$ & $A_{2s}$ & $\tilde{A}_{2p}$ & $\Delta_1$\\ \hline
      $(1,\pm 1)$ & $\quad0.07782\quad$ & $\quad-0.07747i\quad$ & $\quad0.92121\quad$\\ \hline
      $\quad(\frac{1}{\sqrt{2}},\pm \sqrt{\frac{3}{2}})\quad$ & 0.09534 & $-0.08590i$ & 0.92104\\ \hline
      $(0,0)$ & 0.14964 & 0.0 & 0.92181 \\
       \hline
    \end{tabular}
    \caption{Pairing amplitude for various $(\alpha,\beta)$ with $\mu_2=1.5$, $t=0.1$, $V_1=-0.07$, and $160\times160$ k-points is used in momentum summation. $\epsilon_2^{(0)}(\vec{k})=0$ for the top two rows and $\epsilon_2^{(0)}(\vec{k})=|\vec{k}|=\sqrt{k_x^2+k_y^2}$ for the bottom row. Pairing amplitude $A_{2p+ip}$ is related to $\tilde{A}_{2p+ip}$ listed in the table by $A_{2p+ip}=\tilde{A}_{2p+ip}(\alpha\sin(k_x)+i\beta\sin(k_y))$.}
    \label{tablea1}
\end{table}
\begin{table}[t]
\begin{tabular}[b]{|c||c|c|c|}\hline
      $(\alpha,\beta)$ & $A_{2s}$ & $\tilde{A}_{2p}$ & $\Delta_1$\\ \hline
      $(1,\pm 1)$ & $\quad0.02927\quad$ & $\quad-0.00988i\quad$ & $\quad0.92093\quad$\\ \hline
      $\quad(\frac{1}{\sqrt{2}},\pm \sqrt{\frac{3}{2}})\quad$ & 0.03055 & $-0.00860i$ & 0.92085\\ \hline
      $(0,0)$ & 0.30395 & 0.0 & 0.92009 \\
       \hline
\end{tabular}
\caption{Pairing amplitude for various $(\alpha,\beta)$ with $\epsilon_2^{(0)}(\vec{k})=1.8-\big(\cos(k_x)+\cos(k_y)\big)$, $\mu_2=1.5$, $t=0.1$, $V_1=-0.07$, and $160\times160$ k-points is used in momentum summation. Pairing amplitude $A_{2p+ip}$ is related to $\tilde{A}_{2p+ip}$ listed in the table by $A_{2p+ip}=\tilde{A}_{2p+ip}(\alpha\sin(k_x)+i\beta\sin(k_y))$.}
\label{tablea2}
\end{table}

From the numerator of Eq.(\ref{a18}) and Eq.(\ref{a17}), it is clear that the leading pairing amplitude of spin triplet ($\uparrow\uparrow$ or $\downarrow\downarrow$) is of $p\pm ip$ form, and that of spin singlet is of $s$ or $d$ -wave form (depending on the source superconductor, i.e. the symmetry of $\Delta_1(\vec{k})$). This is indeed verified numerically as shown in Table \ref{tablea1} and Table \ref{tablea2}, in which we use $s$-wave superconductor as first layer with dispersion $\bar{\epsilon_1}(\vec{k})=\cos(k_x)+\cos(k_y)+1$. In this computation we have solved the superconducting gap equation self consistently, and before coupling the metallic layer the gap magnitude of the s wave superconductor is $1$ by choosing $V_1=-0.07$.

Eq.(\ref{a18}) also illustrates why the $d$-wave pairing in the superconductor change the pairing orientations as the different signs of $d$ wave pairing gap modifies the relative sign along $k_x$ and $k_y$ directions. Note that there is no $p$-wave pairing in this spin triplet sector $(\uparrow\downarrow+ \downarrow\uparrow)$ but only $p \pm i p$ pairing in the $(\uparrow\uparrow)$ or $(\downarrow\downarrow)$. This has to do with the lack of $\sigma_z$ terms in the model Hamiltonian Eq.(\ref{somat}). By introducing Zeeman field terms\cite{Black} we can also generate the $p$-wave pairing in this $(\uparrow\downarrow+ \downarrow\uparrow)$ spin sector as shown in the discussion of warping terms in Section \ref{TIHTC}.

\section{Comparison between bilayer and trilayer system}\label{AC} 
Here we compare the proximity induced pairing amplitudes of the bottom metallic layer in the bilayer and trilayer system. The bottom metallic layers of both systems have the same linear dispersion $E_N=|\vec{k}|$ and chemical potential $\mu_N=1.5$. For both bilayer and trilayer systems, we choose the same dispersion $E_{SC}=\cos k_x+\cos k_y$ and chemical potential $\mu_{SC}=-1$ in the superconducting layer. The top and bottom Fermi surfaces are chosen to take different forms to highlight the issues of mismatched Fermi surfaces, leading to poor superconducting proximity effect, in general, compared with the matched ones. An intriguing question is whether the middle metallic layer could serve as a good bridging layer to enhance this proximity induced pairing amplitude in the bottom layer, compared with the direct coupling in the bilayer system. 

\begin{figure}[h]
\centering
\subfigure[Bilayer FS]{
\includegraphics[width=.3\columnwidth]{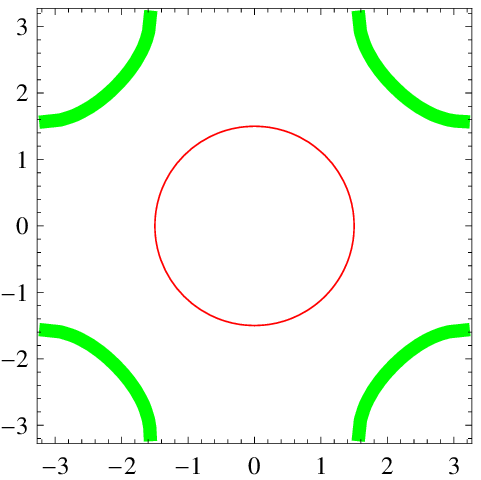}
\label{fig:2FS} } 
\subfigure[Trilayer FS]{
\includegraphics[width=.3\columnwidth]{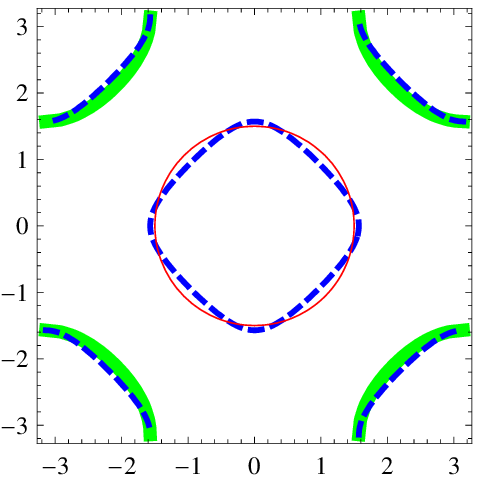}
\label{fig:3FS} }
\subfigure[Trilayer FS2]{
\includegraphics[width=.3\columnwidth]{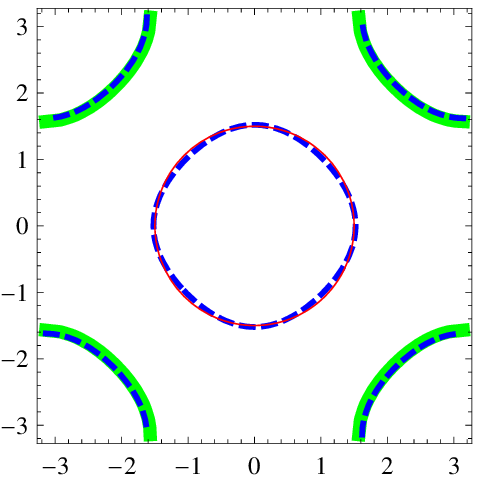}
\label{fig:3FS2} }
\subfigure[Pairing amplitude v.s. normalized hopping]{
\includegraphics[width=.8\columnwidth]{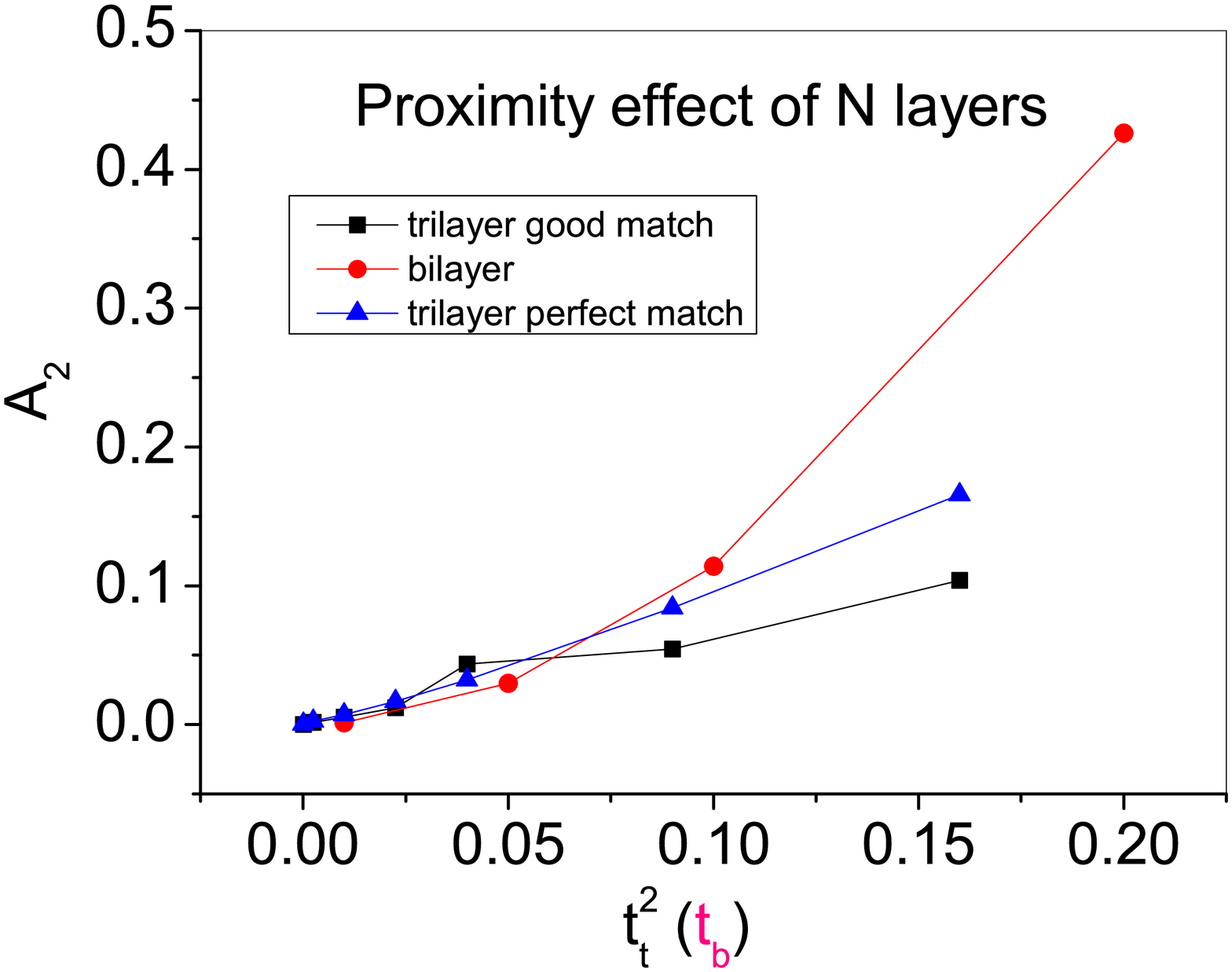}
\label{fig:PA} }
\caption{$E_N=|k|$ and $E_{SC}=\cos k_x+\cos k_y$ with $\mu_N=1.5$ and $\mu_{SC}=-1.0$. $V_{sc1}=0.08$. $40\times40$ k-points. (a). Fermi Surface of the bilayer system (b).Fermi Surface of the trilayer system for $E_{b1}=4\cos k_x\cos k_y+2\cos 2k_x+2\cos 2k_y$ with $\mu_{b1}=0.0$. (c).Fermi Surface of the trilayer system for $E_{b2}=(\cos k_x+\cos k_y -1.05)(\cos k_x +\cos k_y+1.05)$ with $\mu_{b2}=0.0$. (Green for the SC layer, red for the N layer and blue for the bridge layer) (d). Pairing amplitude of the bilayer and trilayer systems.} 
\label{fig:bridge}
\end{figure}
\begin{figure}[h]
\centering
\subfigure[Bilayer FS]{
\includegraphics[width=.45\columnwidth]{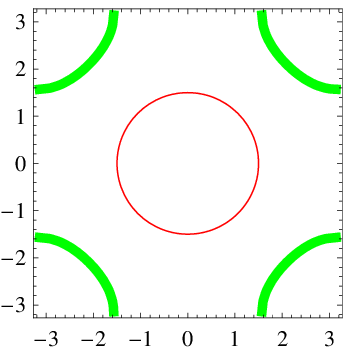}
\label{fig:2FSMis} } 
\subfigure[Trilayer FS]{
\includegraphics[width=.45\columnwidth]{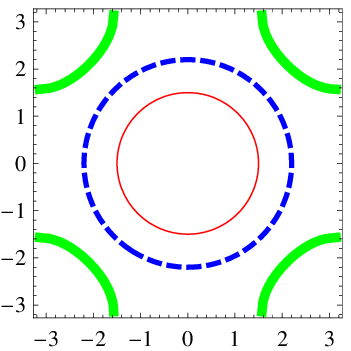}
\label{fig:3FSMis} }
\subfigure[Pairing amplitude v.s. normalized hopping]{
\includegraphics[width=.8\columnwidth]{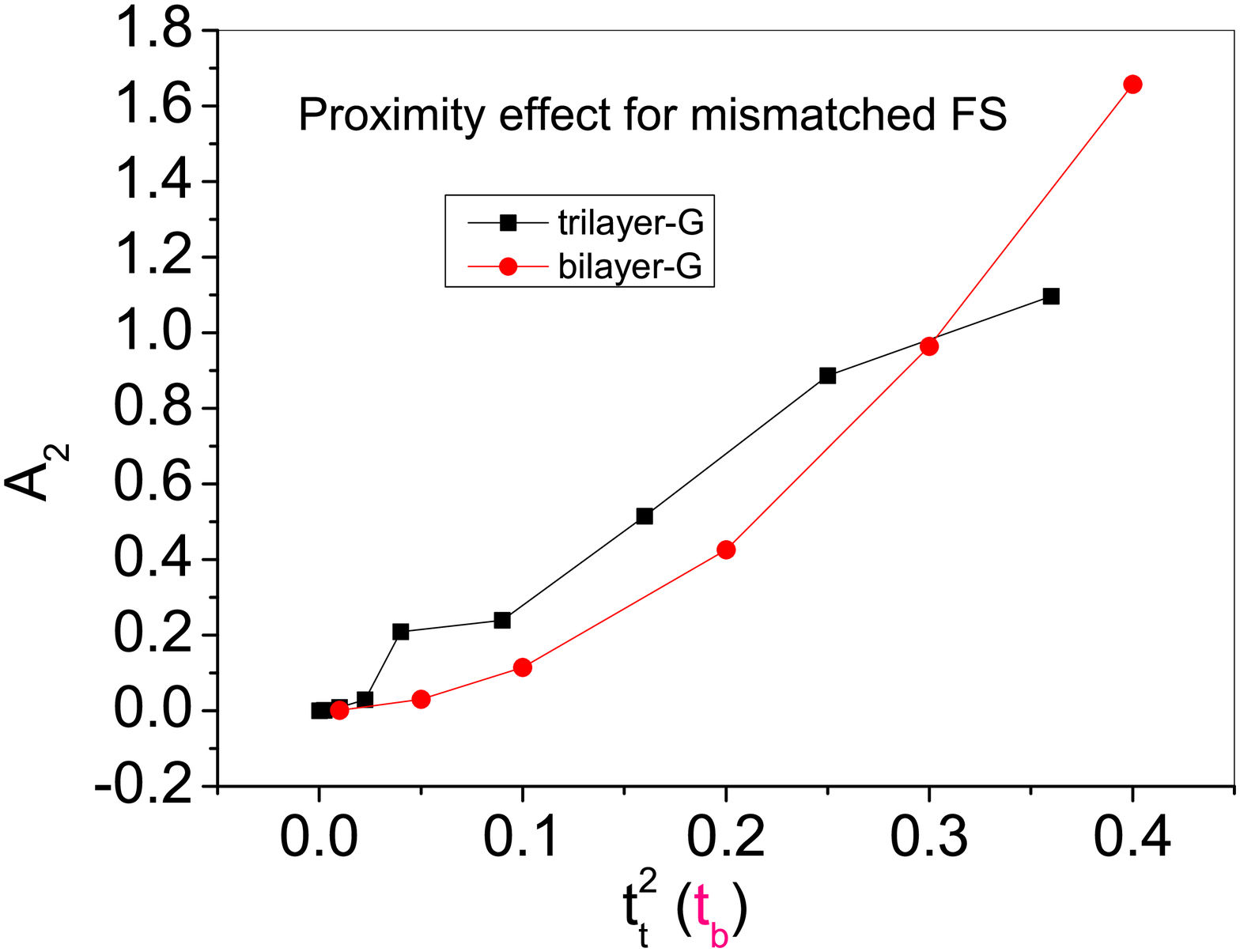}
\label{fig:PAMis} }
\caption{$E_N=|k|$ and $E_{SC}=\cos k_x+\cos k_y$ with $\mu_N=1.5$ and $\mu_{SC}=-1.0$. $V_{sc1}=0.08$. $40\times40$ k-points. (a). Fermi Surface of the bilayer system (b). Fermi Surface of the trilayer system for $E_{b3}=|k|$ with $\mu_{b3}=2.2$. (Green for the SC layer, red for the N layer and blue for the bridge layer) (c). Pairing amplitude of the bilayer and trilayer systems.} 
\label{fig:bridgeMis}
\end{figure} 

The tunneling strength of the bilayer and trilayer systems are labeled by $t_b$ and $t_t$, assuming the tunneling strengths to be identical in the trilayer system for the ease of scaling argument. Compared to the bilayer system, electrons in the trilayer system have to hop twice from the top superconducting layer to the bottom metallic layer. To compare the results of the trilayer systems with those of the bilayer system, we label the effective hopping term of the trilayer system as $t_t^2$ and that of bilayer system as $t_b$ (Here we choose $\Delta_0=1$ so the dimensionless quantity is $(t_t/\Delta_0)^2=t_t^2$. Same for $t_b/\Delta_0=t_b$.). We then compare the pairing amplitudes of $N_1$ and $N_2$ as a function of effective hopping terms $t_b$ and $t_t^2$ respectively. 

To test this bridging layer idea, we choose two special energy dispersions for the middle metallic layer $E_{b1}=4\cos k_x\cos k_y+2\cos 2k_x+2\cos 2k_y$ and $E_{b2}=(\cos k_x+\cos k_y -1.05)(\cos k_x +\cos k_y+1.05)$. The chemical potentials $\mu_{b1}=\mu_{b2}=0$ are chosen such that the Fermi surface covering both that of the top and bottom layer, as shown in Fig.\ref{fig:3FS} and Fig.\ref{fig:3FS2}. We plot the pairing amplitude strengths $A_2$ as a function of the effective hopping terms $t_b$ and $t_t^2$ for bilayer and trilayer systems in Fig.\ref{fig:PA}. As shown in Fig.\ref{fig:PA}, for small effective hopping terms, this choice of bridging layer does enhance the pairing amplitude of the bottom metallic layer. For larger effective hopping terms, however, the bilayer system always has larger pairing amplitudes than the trilayer system regardless of the shape of middle metallic layer Fermi surface.

In contrast to the previous case where the Fermi Surface of the bridging layer matches those of the superconducting and bottom metallic layer, we further study the case where the Fermi Surface of the bridging layer lies between those of the two layers. The same dispersions are chosen for top and bottom layers while the bridging layer has a linear dispersion $E_{b3}=|\vec{k}|$ with the chemical potential $\mu_{b3}>\mu_N$ as shown in Fig.\ref{fig:3FSMis}. The result is shown in Fig.\ref{fig:bridgeMis}. In contrast to the matching of both Fermi surfaces shown in Fig.\ref{fig:bridge}, the enhancement here is relatively large in terms of magnitude and the range of the effective hopping terms. However, at larger tunneling strengths the direct coupling in the bilayer system still gives larger pairing amplitude.  

All of these numerical results can be understood qualitatively from the analytic pairing amplitude derived perturbatively with Hartree approximations in Appendix \ref{AA3}. Under the small tunneling strengths approximation the proximity induced superconductivity pairing amplitude for second and third layer in trilayer system are given in Eq.(\ref{a23}) and Eq.(\ref{a24}), which we rewrite as:
\begin{eqnarray*}
&&\langle c_{\vec{k}_b\uparrow}^{\dagger}c_{-\vec{k}_b\downarrow}^{\dagger}\rangle\simeq \frac{t_{b}^2 \Delta_1(\vec{k})}{2E_1(\vec{k})\big(E_1(\vec{k})^2-\bar{\epsilon}_b(\vec{k})^2\big) }\\
&&\langle c_{\vec{k}_t\uparrow}^{\dagger}c_{-\vec{k}_t\downarrow}^{\dagger}\rangle \simeq \frac{-\Delta_1(\vec{k})t_{t}^4}{2E_1(\vec{k})\big(E_1(\vec{k})^2-\bar{\epsilon}_m(\vec{k})^2\big)\big(E_1(\vec{k})^2-\bar{\epsilon}_b(\vec{k})^2\big)}
\end{eqnarray*}

In above, we replace the tunneling amplitudes by the bilayer and trilayer ones. The form of the second layer pairing amplitude $\langle c_{\vec{k}_2\uparrow}^{\dagger}c_{-\vec{k}_2\downarrow}^{\dagger}\rangle$ in Eq.(\ref{a23}) is the same as the bottom metallic layer pairing amplitude in bilayer system under this Hartree and small tunneling strength(s) approximation, and we relabel the $\langle c_{\vec{k}_2\uparrow}^{\dagger}c_{-\vec{k}_2\downarrow}^{\dagger}\rangle$ by $\langle c_{\vec{k}_b\uparrow}^{\dagger}c_{-\vec{k}_b\downarrow}^{\dagger}\rangle$. Within these approximations $|\langle c_{\vec{k}_t\uparrow}^{\dagger}c_{-\vec{k}_t\downarrow}^{\dagger}\rangle/\langle c_{\vec{k}_b\uparrow}^{\dagger}c_{-\vec{k}_b\downarrow}^{\dagger}\rangle|\simeq (t_{t}^4/t_b^2)/|E_1(\vec{k})^2-\bar{\epsilon}_m(\vec{k})^2|$, which is in general smaller than one even if we set $t_t^2=t_b$. The way to make this ratio larger than one is when $|E_1(\vec{k})|\simeq |\bar{\epsilon}_m(\vec{k})|$, the so called resonant condition in the scattering theory. This condition is made when the middle layer dispersion $\bar{\epsilon}_m(\vec{k})$ are not completely in line with the original one given by $\bar{\epsilon}_1(\vec{k})$, but with the Bogoliubov quasi particle level $E_1(\vec{k})$. This is why the range and magnitude of enhancement shown in Fig.\ref{fig:bridgeMis} is greater than that in Fig.\ref{fig:bridge}. At larger tunneling strength, the shift of energy level by the single particle tunneling becomes more important, and the proximity induced gap actually decreases as discussed in Appendix.\ref{AC}. As the suppression of pairing amplitude in trilayer is twice of the bilayer, the bilayer always has greater pairing amplitude at larger tunneling strengths as shown in Fig.\ref{fig:bridge} and Fig.\ref{fig:bridgeMis}. 

Knowing that adding an additional middle metallic layer in general does not help improve the proximity effect, we discuss the issues of lattice mismatch, which is quite general, but difficult to compute for incommensurate lattice ratios, for interface between different materials in the next section.

\section{Lattice mismatch issues}\label{AD}
For different materials, the lattice sizes and shapes are usually different. Here we focus our discussions on the cases of square lattice, and the lattice mismatch discussed here means different lattice lengths. We compare the obtained pairing amplitude $A_2$ in the metallic layer and the self consistent gap magnitude $\Delta_1$ of the superconducting layer. For matched Fermi surfaces and lattice sizes we choose $\epsilon_1(\vec{k})=-\epsilon_2(\vec{k})=\cos (k_x)+\cos(k_y)$, $\mu_1=-\mu_2=-1.0$, with $1$ and $2$ denoting superconducting and metallic layer and tunneling amplitude $t=0.1$. For mismatched lattices we choose $\epsilon_1(\vec{k})=\cos(3 k_x)+\cos(3 k_y)$, $\mu_1=-\mu_2=-1.0$, $-\epsilon_2(\vec{k})=\cos(4 k_x)+\cos(4 k_y)$, $t=0.1$ and $V_1=-0.07$. Similar choice is done for the mismatched Fermi surface with parameters chosen as $\epsilon_1(\vec{k})=\epsilon_2(\vec{k})=\cos(k_x)+\cos(k_y)$, $\mu_1=-\mu_2=-1.0$, $t=0.1$, and $V_1=-0.07$. The results are listed in Table \ref{tablemis}.

  It is clear that for originally matched Fermi surfaces, the factor of mismatched lattice makes less overlapping region of Fermi levels. Therefore we shall expect the decrease in the proximity effect. For the originally mismatched Fermi surfaces, this lattice mismatch factor actually increases the percentage of overlapping Fermi levels, and possibly enhance the proximity effect compared with original identical lattices. These intuitions are consistent with what we obtained in the numerical results shown in Table \ref{tablemis}. 

\begin{table}[t]
\begin{tabular}[b]{|c||c|c|c|c|}\hline
      Cases & FSM, LM & FSM, LmM & FSmM, LM & FSmM, LmM \\ \hline
      $A_2$ & $-0.42214$ & $ -0.27422$ & $-0.13747$ & $-0.25523$ \\ \hline
      $\Delta_1$ & $0.90649$ & $0.91302$ & $0.91354$ & $0.91299$ \\
       \hline
\end{tabular}
\caption{Terminologies used in this table: FSM = Fermi surface matched; LM = Lattice matched; FSmM = Fermi surface mismatched; LmM = Lattice mismatched. For lattice mismatch cases we choose the ratio of lattice size in superconductor/metal as $3/4$. All other model parameters are specified in the text of Appendix.\ref{AD}}
\label{tablemis}
\end{table}

\bibliographystyle{apsrev4-1}
\bibliography{proximity}

\begin{thebibliography}{31}%
\makeatletter
\providecommand \@ifxundefined [1]{%
 \@ifx{#1\undefined}
}%
\providecommand \@ifnum [1]{%
 \ifnum #1\expandafter \@firstoftwo
 \else \expandafter \@secondoftwo
 \fi
}%
\providecommand \@ifx [1]{%
 \ifx #1\expandafter \@firstoftwo
 \else \expandafter \@secondoftwo
 \fi
}%
\providecommand \natexlab [1]{#1}%
\providecommand \enquote  [1]{``#1''}%
\providecommand \bibnamefont  [1]{#1}%
\providecommand \bibfnamefont [1]{#1}%
\providecommand \citenamefont [1]{#1}%
\providecommand \href@noop [0]{\@secondoftwo}%
\providecommand \href [0]{\begingroup \@sanitize@url \@href}%
\providecommand \@href[1]{\@@startlink{#1}\@@href}%
\providecommand \@@href[1]{\endgroup#1\@@endlink}%
\providecommand \@sanitize@url [0]{\catcode `\\12\catcode `\$12\catcode
  `\&12\catcode `\#12\catcode `\^12\catcode `\_12\catcode `\%12\relax}%
\providecommand \@@startlink[1]{}%
\providecommand \@@endlink[0]{}%
\providecommand \url  [0]{\begingroup\@sanitize@url \@url }%
\providecommand \@url [1]{\endgroup\@href {#1}{\urlprefix }}%
\providecommand \urlprefix  [0]{URL }%
\providecommand \Eprint [0]{\href }%
\providecommand \doibase [0]{http://dx.doi.org/}%
\providecommand \selectlanguage [0]{\@gobble}%
\providecommand \bibinfo  [0]{\@secondoftwo}%
\providecommand \bibfield  [0]{\@secondoftwo}%
\providecommand \translation [1]{[#1]}%
\providecommand \BibitemOpen [0]{}%
\providecommand \bibitemStop [0]{}%
\providecommand \bibitemNoStop [0]{.\EOS\space}%
\providecommand \EOS [0]{\spacefactor3000\relax}%
\providecommand \BibitemShut  [1]{\csname bibitem#1\endcsname}%
\let\auto@bib@innerbib\@empty
\bibitem [{\citenamefont {Meissner}(1960)}]{Meissner1960}%
  \BibitemOpen
  \bibfield  {author} {\bibinfo {author} {\bibfnamefont {H.}~\bibnamefont
  {Meissner}},\ }\href@noop {} {\bibfield  {journal} {\bibinfo  {journal}
  {Phys. Rev.}\ }\textbf {\bibinfo {volume} {117}},\ \bibinfo {pages} {672}
  (\bibinfo {year} {1960})}\BibitemShut {NoStop}%
\bibitem [{\citenamefont {Steiner}\ and\ \citenamefont
  {Ziemann}(2006)}]{Steiner2006}%
  \BibitemOpen
  \bibfield  {author} {\bibinfo {author} {\bibfnamefont {R.}~\bibnamefont
  {Steiner}}\ and\ \bibinfo {author} {\bibfnamefont {P.}~\bibnamefont
  {Ziemann}},\ }\href@noop {} {\bibfield  {journal} {\bibinfo  {journal}
  {Physical Review B}\ }\textbf {\bibinfo {volume} {74}},\ \bibinfo {pages}
  {094504} (\bibinfo {year} {2006})}\BibitemShut {NoStop}%
\bibitem [{\citenamefont {Cirillo}\ \emph {et~al.}(2005)\citenamefont
  {Cirillo}, \citenamefont {Prischepa}, \citenamefont {Salvato}, \citenamefont
  {Attanasio}, \citenamefont {Hesselberth},\ and\ \citenamefont
  {Aarts}}]{Cirillo2005}%
  \BibitemOpen
  \bibfield  {author} {\bibinfo {author} {\bibfnamefont {C.}~\bibnamefont
  {Cirillo}}, \bibinfo {author} {\bibfnamefont {S.}~\bibnamefont {Prischepa}},
  \bibinfo {author} {\bibfnamefont {M.}~\bibnamefont {Salvato}}, \bibinfo
  {author} {\bibfnamefont {C.}~\bibnamefont {Attanasio}}, \bibinfo {author}
  {\bibfnamefont {M.}~\bibnamefont {Hesselberth}}, \ and\ \bibinfo {author}
  {\bibfnamefont {J.}~\bibnamefont {Aarts}},\ }\href@noop {} {\bibfield
  {journal} {\bibinfo  {journal} {Physical Review B}\ }\textbf {\bibinfo
  {volume} {72}},\ \bibinfo {pages} {144511} (\bibinfo {year}
  {2005})}\BibitemShut {NoStop}%
\bibitem [{\citenamefont {Tesauro}\ \emph {et~al.}(2005)\citenamefont
  {Tesauro}, \citenamefont {Aurigemma}, \citenamefont {Cirillo}, \citenamefont
  {Prischepa}, \citenamefont {Salvato},\ and\ \citenamefont
  {Attanasio}}]{Tesauro2005}%
  \BibitemOpen
  \bibfield  {author} {\bibinfo {author} {\bibfnamefont {A.}~\bibnamefont
  {Tesauro}}, \bibinfo {author} {\bibfnamefont {A.}~\bibnamefont {Aurigemma}},
  \bibinfo {author} {\bibfnamefont {C.}~\bibnamefont {Cirillo}}, \bibinfo
  {author} {\bibfnamefont {S.~L.}\ \bibnamefont {Prischepa}}, \bibinfo {author}
  {\bibfnamefont {M.}~\bibnamefont {Salvato}}, \ and\ \bibinfo {author}
  {\bibfnamefont {C.}~\bibnamefont {Attanasio}},\ }\href@noop {} {\bibfield
  {journal} {\bibinfo  {journal} {Superconductor Science and Technology}\
  }\textbf {\bibinfo {volume} {18}},\ \bibinfo {pages} {1} (\bibinfo {year}
  {2005})}\BibitemShut {NoStop}%
\bibitem [{\citenamefont {Bergeret}\ \emph {et~al.}(2001)\citenamefont
  {Bergeret}, \citenamefont {Volkov},\ and\ \citenamefont
  {Efetov}}]{Bergeret2001}%
  \BibitemOpen
  \bibfield  {author} {\bibinfo {author} {\bibfnamefont {F.~S.}\ \bibnamefont
  {Bergeret}}, \bibinfo {author} {\bibfnamefont {A.~F.}\ \bibnamefont
  {Volkov}}, \ and\ \bibinfo {author} {\bibfnamefont {K.~B.}\ \bibnamefont
  {Efetov}},\ }\href {\doibase 10.1103/PhysRevLett.86.4096} {\bibfield
  {journal} {\bibinfo  {journal} {Physical Review Letters}\ }\textbf {\bibinfo
  {volume} {86}},\ \bibinfo {pages} {4096} (\bibinfo {year}
  {2001})}\BibitemShut {NoStop}%
\bibitem [{\citenamefont {Keizer}\ \emph {et~al.}(2006)\citenamefont {Keizer},
  \citenamefont {Goennenwein}, \citenamefont {Klapwijk}, \citenamefont {Miao},
  \citenamefont {Xiao},\ and\ \citenamefont {Gupta}}]{Keizer2006}%
  \BibitemOpen
  \bibfield  {author} {\bibinfo {author} {\bibfnamefont {R.~S.}\ \bibnamefont
  {Keizer}}, \bibinfo {author} {\bibfnamefont {S.~T.~B.}\ \bibnamefont
  {Goennenwein}}, \bibinfo {author} {\bibfnamefont {T.~M.}\ \bibnamefont
  {Klapwijk}}, \bibinfo {author} {\bibfnamefont {G.}~\bibnamefont {Miao}},
  \bibinfo {author} {\bibfnamefont {G.}~\bibnamefont {Xiao}}, \ and\ \bibinfo
  {author} {\bibfnamefont {a.}~\bibnamefont {Gupta}},\ }\href@noop {}
  {\bibfield  {journal} {\bibinfo  {journal} {Nature}\ }\textbf {\bibinfo
  {volume} {439}},\ \bibinfo {pages} {825} (\bibinfo {year}
  {2006})}\BibitemShut {NoStop}%
\bibitem [{\citenamefont {Fu}\ and\ \citenamefont {Kane}(2008)}]{Fu2008}%
  \BibitemOpen
  \bibfield  {author} {\bibinfo {author} {\bibfnamefont {L.}~\bibnamefont
  {Fu}}\ and\ \bibinfo {author} {\bibfnamefont {C.~L.}\ \bibnamefont {Kane}},\
  }\href@noop {} {\bibfield  {journal} {\bibinfo  {journal} {Phys. Rev. Lett.}\
  }\textbf {\bibinfo {volume} {100}},\ \bibinfo {pages} {096407} (\bibinfo
  {year} {2008})}\BibitemShut {NoStop}%
\bibitem [{\citenamefont {Lee}\ \emph {et~al.}(2014)\citenamefont {Lee},
  \citenamefont {Vaezi}, \citenamefont {Fischer},\ and\ \citenamefont
  {Kim}}]{Kim2014}%
  \BibitemOpen
  \bibfield  {author} {\bibinfo {author} {\bibfnamefont {K.}~\bibnamefont
  {Lee}}, \bibinfo {author} {\bibfnamefont {A.}~\bibnamefont {Vaezi}}, \bibinfo
  {author} {\bibfnamefont {M.~H.}\ \bibnamefont {Fischer}}, \ and\ \bibinfo
  {author} {\bibfnamefont {E.~A.}\ \bibnamefont {Kim}},\ }\href@noop {}
  {\bibfield  {journal} {\bibinfo  {journal} {Phys. Rev. B}\ }\textbf {\bibinfo
  {volume} {90}},\ \bibinfo {pages} {214510} (\bibinfo {year}
  {2014})}\BibitemShut {NoStop}%
\bibitem [{\citenamefont {Black-Schaffer}\ and\ \citenamefont
  {Balatsky}(2013)}]{Black-Schaffer2013}%
  \BibitemOpen
  \bibfield  {author} {\bibinfo {author} {\bibfnamefont {A.~M.}\ \bibnamefont
  {Black-Schaffer}}\ and\ \bibinfo {author} {\bibfnamefont {A.~V.}\
  \bibnamefont {Balatsky}},\ }\href@noop {} {\bibfield  {journal} {\bibinfo
  {journal} {Phys. Rev. B}\ }\textbf {\bibinfo {volume} {87}},\ \bibinfo
  {pages} {220506} (\bibinfo {year} {2013})}\BibitemShut {NoStop}%
\bibitem [{\citenamefont {Takane}\ and\ \citenamefont
  {Ando}(2014)}]{Takane2014}%
  \BibitemOpen
  \bibfield  {author} {\bibinfo {author} {\bibfnamefont {Y.}~\bibnamefont
  {Takane}}\ and\ \bibinfo {author} {\bibfnamefont {R.}~\bibnamefont {Ando}},\
  }\href@noop {} {\bibfield  {journal} {\bibinfo  {journal} {J. Phys. Soc.
  Jpn}\ }\textbf {\bibinfo {volume} {83}},\ \bibinfo {pages} {014706} (\bibinfo
  {year} {2014})}\BibitemShut {NoStop}%
\bibitem [{\citenamefont {Linder}\ \emph
  {et~al.}(2010{\natexlab{a}})\citenamefont {Linder}, \citenamefont {Tanaka},
  \citenamefont {Yokoyama}, \citenamefont {Sudbo},\ and\ \citenamefont
  {Nagaosa}}]{Linder2010}%
  \BibitemOpen
  \bibfield  {author} {\bibinfo {author} {\bibfnamefont {J.}~\bibnamefont
  {Linder}}, \bibinfo {author} {\bibfnamefont {Y.}~\bibnamefont {Tanaka}},
  \bibinfo {author} {\bibfnamefont {T.}~\bibnamefont {Yokoyama}}, \bibinfo
  {author} {\bibfnamefont {A.}~\bibnamefont {Sudbo}}, \ and\ \bibinfo {author}
  {\bibfnamefont {N.}~\bibnamefont {Nagaosa}},\ }\href@noop {} {\bibfield
  {journal} {\bibinfo  {journal} {Phys. Rev. Lett.}\ }\textbf {\bibinfo
  {volume} {104}},\ \bibinfo {pages} {067001} (\bibinfo {year}
  {2010}{\natexlab{a}})}\BibitemShut {NoStop}%
\bibitem [{\citenamefont {Linder}\ \emph
  {et~al.}(2010{\natexlab{b}})\citenamefont {Linder}, \citenamefont {Tanaka},
  \citenamefont {Yokoyama}, \citenamefont {Sudbo},\ and\ \citenamefont
  {Nagaosa}}]{Linder2010a}%
  \BibitemOpen
  \bibfield  {author} {\bibinfo {author} {\bibfnamefont {J.}~\bibnamefont
  {Linder}}, \bibinfo {author} {\bibfnamefont {Y.}~\bibnamefont {Tanaka}},
  \bibinfo {author} {\bibfnamefont {T.}~\bibnamefont {Yokoyama}}, \bibinfo
  {author} {\bibfnamefont {A.}~\bibnamefont {Sudbo}}, \ and\ \bibinfo {author}
  {\bibfnamefont {N.}~\bibnamefont {Nagaosa}},\ }\href@noop {} {\bibfield
  {journal} {\bibinfo  {journal} {Phys. Rev. B.}\ }\textbf {\bibinfo {volume}
  {81}},\ \bibinfo {pages} {184525} (\bibinfo {year}
  {2010}{\natexlab{b}})}\BibitemShut {NoStop}%
\bibitem [{\citenamefont {Zareapour}\ \emph {et~al.}(2012)\citenamefont
  {Zareapour}, \citenamefont {Hayat}, \citenamefont {Zhao}, \citenamefont
  {Kreshchuk}, \citenamefont {Jain}, \citenamefont {Kwok}, \citenamefont {Lee},
  \citenamefont {Cheong}, \citenamefont {Xu}, \citenamefont {Yang},
  \citenamefont {Gu}, \citenamefont {Jia}, \citenamefont {Cava},\ and\
  \citenamefont {Burch}}]{Parisa2012}%
  \BibitemOpen
  \bibfield  {author} {\bibinfo {author} {\bibfnamefont {P.}~\bibnamefont
  {Zareapour}}, \bibinfo {author} {\bibfnamefont {A.}~\bibnamefont {Hayat}},
  \bibinfo {author} {\bibfnamefont {S.~Y.~F.}\ \bibnamefont {Zhao}}, \bibinfo
  {author} {\bibfnamefont {M.}~\bibnamefont {Kreshchuk}}, \bibinfo {author}
  {\bibfnamefont {A.}~\bibnamefont {Jain}}, \bibinfo {author} {\bibfnamefont
  {D.~C.}\ \bibnamefont {Kwok}}, \bibinfo {author} {\bibfnamefont
  {N.}~\bibnamefont {Lee}}, \bibinfo {author} {\bibfnamefont {S.-W.}\
  \bibnamefont {Cheong}}, \bibinfo {author} {\bibfnamefont {Z.}~\bibnamefont
  {Xu}}, \bibinfo {author} {\bibfnamefont {A.}~\bibnamefont {Yang}}, \bibinfo
  {author} {\bibfnamefont {G.}~\bibnamefont {Gu}}, \bibinfo {author}
  {\bibfnamefont {S.}~\bibnamefont {Jia}}, \bibinfo {author} {\bibfnamefont
  {R.~J.}\ \bibnamefont {Cava}}, \ and\ \bibinfo {author} {\bibfnamefont
  {K.~S.}\ \bibnamefont {Burch}},\ }\href@noop {} {\bibfield  {journal}
  {\bibinfo  {journal} {Nat Commun}\ }\textbf {\bibinfo {volume} {3}},\
  \bibinfo {pages} {1056} (\bibinfo {year} {2012})}\BibitemShut {NoStop}%
\bibitem [{\citenamefont {Yilmaz}\ \emph {et~al.}(2014)\citenamefont {Yilmaz},
  \citenamefont {Pletikosi\'{c}}, \citenamefont {Weber}, \citenamefont
  {Sadowski}, \citenamefont {Gu}, \citenamefont {Caruso}, \citenamefont
  {Sinkovic},\ and\ \citenamefont {Valla}}]{Yilmaz2014}%
  \BibitemOpen
  \bibfield  {author} {\bibinfo {author} {\bibfnamefont {T.}~\bibnamefont
  {Yilmaz}}, \bibinfo {author} {\bibfnamefont {I.}~\bibnamefont
  {Pletikosi\'{c}}}, \bibinfo {author} {\bibfnamefont {A.~P.}\ \bibnamefont
  {Weber}}, \bibinfo {author} {\bibfnamefont {J.~T.}\ \bibnamefont {Sadowski}},
  \bibinfo {author} {\bibfnamefont {G.~D.}\ \bibnamefont {Gu}}, \bibinfo
  {author} {\bibfnamefont {A.~N.}\ \bibnamefont {Caruso}}, \bibinfo {author}
  {\bibfnamefont {B.}~\bibnamefont {Sinkovic}}, \ and\ \bibinfo {author}
  {\bibfnamefont {T.}~\bibnamefont {Valla}},\ }\href@noop {} {\bibfield
  {journal} {\bibinfo  {journal} {Phys. Rev. Lett.}\ }\textbf {\bibinfo
  {volume} {113}},\ \bibinfo {pages} {067003} (\bibinfo {year}
  {2014})}\BibitemShut {NoStop}%
\bibitem [{\citenamefont {Xu}\ \emph {et~al.}(2014)\citenamefont {Xu},
  \citenamefont {Liu}, \citenamefont {Richardella}, \citenamefont {Belopolski},
  \citenamefont {Alidoust}, \citenamefont {Neupane}, \citenamefont {Bian},
  \citenamefont {Samarth},\ and\ \citenamefont {Hasan}}]{Xu2014}%
  \BibitemOpen
  \bibfield  {author} {\bibinfo {author} {\bibfnamefont {S.-Y.}\ \bibnamefont
  {Xu}}, \bibinfo {author} {\bibfnamefont {C.}~\bibnamefont {Liu}}, \bibinfo
  {author} {\bibfnamefont {A.}~\bibnamefont {Richardella}}, \bibinfo {author}
  {\bibfnamefont {I.}~\bibnamefont {Belopolski}}, \bibinfo {author}
  {\bibfnamefont {N.}~\bibnamefont {Alidoust}}, \bibinfo {author}
  {\bibfnamefont {M.}~\bibnamefont {Neupane}}, \bibinfo {author} {\bibfnamefont
  {G.}~\bibnamefont {Bian}}, \bibinfo {author} {\bibfnamefont {N.}~\bibnamefont
  {Samarth}}, \ and\ \bibinfo {author} {\bibfnamefont {M.~Z.}\ \bibnamefont
  {Hasan}},\ }\href {\doibase 10.1103/PhysRevB.90.085128} {\bibfield  {journal}
  {\bibinfo  {journal} {Phys. Rev. B}\ }\textbf {\bibinfo {volume} {90}},\
  \bibinfo {pages} {085128} (\bibinfo {year} {2014})}\BibitemShut {NoStop}%
\bibitem [{\citenamefont {Wang}\ \emph {et~al.}(2013)\citenamefont {Wang},
  \citenamefont {Ding}, \citenamefont {Fedorov}, \citenamefont {Yao},
  \citenamefont {Li}, \citenamefont {Lv}, \citenamefont {Zhao}, \citenamefont
  {Zhang}, \citenamefont {Xu}, \citenamefont {Schneeloch}, \citenamefont
  {Zhong}, \citenamefont {Ji}, \citenamefont {Wang}, \citenamefont {He},
  \citenamefont {Ma}, \citenamefont {Gu}, \citenamefont {Yao}, \citenamefont
  {Xue}, \citenamefont {Chen},\ and\ \citenamefont {Zhou}}]{Wang2013}%
  \BibitemOpen
  \bibfield  {author} {\bibinfo {author} {\bibfnamefont {E.}~\bibnamefont
  {Wang}}, \bibinfo {author} {\bibfnamefont {H.}~\bibnamefont {Ding}}, \bibinfo
  {author} {\bibfnamefont {A.~V.}\ \bibnamefont {Fedorov}}, \bibinfo {author}
  {\bibfnamefont {W.}~\bibnamefont {Yao}}, \bibinfo {author} {\bibfnamefont
  {Z.}~\bibnamefont {Li}}, \bibinfo {author} {\bibfnamefont {Y.-F.}\
  \bibnamefont {Lv}}, \bibinfo {author} {\bibfnamefont {K.}~\bibnamefont
  {Zhao}}, \bibinfo {author} {\bibfnamefont {L.-G.}\ \bibnamefont {Zhang}},
  \bibinfo {author} {\bibfnamefont {Z.}~\bibnamefont {Xu}}, \bibinfo {author}
  {\bibfnamefont {J.}~\bibnamefont {Schneeloch}}, \bibinfo {author}
  {\bibfnamefont {R.}~\bibnamefont {Zhong}}, \bibinfo {author} {\bibfnamefont
  {S.-H.}\ \bibnamefont {Ji}}, \bibinfo {author} {\bibfnamefont
  {L.}~\bibnamefont {Wang}}, \bibinfo {author} {\bibfnamefont {K.}~\bibnamefont
  {He}}, \bibinfo {author} {\bibfnamefont {X.}~\bibnamefont {Ma}}, \bibinfo
  {author} {\bibfnamefont {G.}~\bibnamefont {Gu}}, \bibinfo {author}
  {\bibfnamefont {H.}~\bibnamefont {Yao}}, \bibinfo {author} {\bibfnamefont
  {Q.-K.}\ \bibnamefont {Xue}}, \bibinfo {author} {\bibfnamefont
  {X.}~\bibnamefont {Chen}}, \ and\ \bibinfo {author} {\bibfnamefont
  {S.}~\bibnamefont {Zhou}},\ }\href@noop {} {\bibfield  {journal} {\bibinfo
  {journal} {Nat. Phys.}\ }\textbf {\bibinfo {volume} {9}},\ \bibinfo {pages}
  {621} (\bibinfo {year} {2013})}\BibitemShut {NoStop}%
\bibitem [{\citenamefont {McMillan}(1968)}]{McMillan1968}%
  \BibitemOpen
  \bibfield  {author} {\bibinfo {author} {\bibfnamefont {W.~L.}\ \bibnamefont
  {McMillan}},\ }\href@noop {} {\bibfield  {journal} {\bibinfo  {journal}
  {Phys. Rev.}\ }\textbf {\bibinfo {volume} {175}},\ \bibinfo {pages} {537}
  (\bibinfo {year} {1968})}\BibitemShut {NoStop}%
\bibitem [{Note1()}]{Note1}%
  \BibitemOpen
  \bibinfo {note} {Here we treat the BSCCO as the bulk, and its
  superconductivity is not influenced by the contact surface with $\protect
  \text {Bi}_2\protect \text {Se}_3$. More rigorously this interface gap should
  be evaluated self consistently, and the gap magnitude of BSCCO at the
  interface would be smaller than its bulk value. Considering this would
  further decrease the proximity induced gap at the $\protect \text
  {Bi}_2\protect \text {Se}_3$. Thus the gap obtained in the main text should
  be viewed as the theoretical upper bound.}\BibitemShut {Stop}%
\bibitem [{\citenamefont {Li}\ \emph {et~al.}(2015)\citenamefont {Li},
  \citenamefont {Chan},\ and\ \citenamefont {Yao}}]{Yao2015}%
  \BibitemOpen
  \bibfield  {author} {\bibinfo {author} {\bibfnamefont {Z.-X.}\ \bibnamefont
  {Li}}, \bibinfo {author} {\bibfnamefont {C.}~\bibnamefont {Chan}}, \ and\
  \bibinfo {author} {\bibfnamefont {H.}~\bibnamefont {Yao}},\ }\href@noop {}
  {\bibfield  {journal} {\bibinfo  {journal} {Phys. Rev. B.}\ }\textbf
  {\bibinfo {volume} {91}},\ \bibinfo {pages} {235143} (\bibinfo {year}
  {2015})}\BibitemShut {NoStop}%
\bibitem [{\citenamefont {Liu}\ \emph {et~al.}(2010)\citenamefont {Liu},
  \citenamefont {Qi}, \citenamefont {Zhang}, \citenamefont {Dai}, \citenamefont
  {Fang},\ and\ \citenamefont {Zhang}}]{Liu2010}%
  \BibitemOpen
  \bibfield  {author} {\bibinfo {author} {\bibfnamefont {C.~X.}\ \bibnamefont
  {Liu}}, \bibinfo {author} {\bibfnamefont {X.~L.}\ \bibnamefont {Qi}},
  \bibinfo {author} {\bibfnamefont {H.}~\bibnamefont {Zhang}}, \bibinfo
  {author} {\bibfnamefont {X.}~\bibnamefont {Dai}}, \bibinfo {author}
  {\bibfnamefont {Z.}~\bibnamefont {Fang}}, \ and\ \bibinfo {author}
  {\bibfnamefont {S.~C.}\ \bibnamefont {Zhang}},\ }\href {\doibase
  10.1103/PhysRevB.82.045122} {\bibfield  {journal} {\bibinfo  {journal} {Phys.
  Rev. B.}\ }\textbf {\bibinfo {volume} {82}},\ \bibinfo {pages} {045122}
  (\bibinfo {year} {2010})}\BibitemShut {NoStop}%
\bibitem [{\citenamefont {Zhang}\ \emph {et~al.}(2010)\citenamefont {Zhang},
  \citenamefont {Yu}, \citenamefont {Zhang}, \citenamefont {Dai},\ and\
  \citenamefont {Fang}}]{Zhang2010}%
  \BibitemOpen
  \bibfield  {author} {\bibinfo {author} {\bibfnamefont {W.}~\bibnamefont
  {Zhang}}, \bibinfo {author} {\bibfnamefont {R.}~\bibnamefont {Yu}}, \bibinfo
  {author} {\bibfnamefont {H.~J.}\ \bibnamefont {Zhang}}, \bibinfo {author}
  {\bibfnamefont {X.}~\bibnamefont {Dai}}, \ and\ \bibinfo {author}
  {\bibfnamefont {Z.}~\bibnamefont {Fang}},\ }\href@noop {} {\bibfield
  {journal} {\bibinfo  {journal} {New J. Phys.}\ }\textbf {\bibinfo {volume}
  {12}},\ \bibinfo {pages} {065013} (\bibinfo {year} {2010})}\BibitemShut
  {NoStop}%
\bibitem [{\citenamefont {Norman}\ \emph {et~al.}(1995)\citenamefont {Norman},
  \citenamefont {Randeria}, \citenamefont {Ding},\ and\ \citenamefont
  {Campuzano}}]{Norman1995}%
  \BibitemOpen
  \bibfield  {author} {\bibinfo {author} {\bibfnamefont {M.~R.}\ \bibnamefont
  {Norman}}, \bibinfo {author} {\bibfnamefont {M.}~\bibnamefont {Randeria}},
  \bibinfo {author} {\bibfnamefont {H.}~\bibnamefont {Ding}}, \ and\ \bibinfo
  {author} {\bibfnamefont {J.~C.}\ \bibnamefont {Campuzano}},\ }\href@noop {}
  {\bibfield  {journal} {\bibinfo  {journal} {Phys. Rev. B.}\ }\textbf
  {\bibinfo {volume} {52}},\ \bibinfo {pages} {615} (\bibinfo {year}
  {1995})}\BibitemShut {NoStop}%
\bibitem [{\citenamefont {Parhizgar}\ and\ \citenamefont
  {Black-Schaffer}(2014)}]{Black}%
  \BibitemOpen
  \bibfield  {author} {\bibinfo {author} {\bibfnamefont {F.}~\bibnamefont
  {Parhizgar}}\ and\ \bibinfo {author} {\bibfnamefont {A.~M.}\ \bibnamefont
  {Black-Schaffer}},\ }\href@noop {} {\bibfield  {journal} {\bibinfo  {journal}
  {Phys. Rev. B}\ }\textbf {\bibinfo {volume} {90}},\ \bibinfo {pages} {184517}
  (\bibinfo {year} {2014})}\BibitemShut {NoStop}%
\bibitem [{Note2()}]{Note2}%
  \BibitemOpen
  \bibinfo {note} {The $d$ wave gap reads out from the DOS might be larger than
  its actual value as the normal state bands also mix with the spectral weight
  of quasiparticles.}\BibitemShut {Stop}%
\bibitem [{Note3()}]{Note3}%
  \BibitemOpen
  \bibinfo {note} {The inclusion of bulk band decreases the pairing gap at the
  contact surface, but it helps enlarge the effective tunneling term (from
  insulating to metallic) between the contact surface and the surface measured
  in ARPES. Large interfacial tunneling term helps maintain the magnitude of
  the pairing gap measured in ARPES as shown in model discussion in
  Appendix.\ref {AC}. Here the focus is put on whether it helps enlarge the
  proximity induced gap at the contact surface.}\BibitemShut {Stop}%
\bibitem [{Note4()}]{Note4}%
  \BibitemOpen
  \bibinfo {note} {$\protect \text {Bi}_2\protect \text {Se}_3$ is not
  superconducting at ambient pressure, but the high pressure leading to
  structural change does lead to superconductivity. See P. P. Kong et. al., J.
  Phys.: Condens. Matter 25, 362204 (2013). Also in $\protect \text
  {Bi}_2\protect \text {Se}_3$ made by pulsed laser deposition the $\protect
  \text {Bi}_2\protect \text {Se}_3$ becomes superconducting due to proximity
  effect from the superconducting $\protect \text {Bi}$ islands grown on its
  surface. See P. H. Le et. al., APL Materials 2, 096105 (2014).}\BibitemShut
  {Stop}%
\bibitem [{Note5()}]{Note5}%
  \BibitemOpen
  \bibinfo {note} {This critical value depends on the tight binding model of
  BSCCO we use, but the physics of hybrid gap, or coherence gap mixing with
  single particle gap, mentioned in the main text exists for all different
  models. For Ref.\protect \rev@citealpnum {Norman1995} it is $t/h=2$ and for
  the tight binding model used in Ref.\protect \rev@citealpnum {Annett2002} it
  is $t/h=1.3$. For tight binding model used in Ref.\protect \rev@citealpnum
  {Hoogenboom,Fink2003} the critical $t/h$ is around $3$. This value mainly
  depends on the energy dispersion of BSCCO at $\Gamma $ point in momentum
  space.}\BibitemShut {Stop}%
\bibitem [{\citenamefont {Kamar}\ and\ \citenamefont
  {Vidhyadhiraja}(2015)}]{Kamar}%
  \BibitemOpen
  \bibfield  {author} {\bibinfo {author} {\bibfnamefont {N.~A.}\ \bibnamefont
  {Kamar}}\ and\ \bibinfo {author} {\bibfnamefont {N.~S.}\ \bibnamefont
  {Vidhyadhiraja}},\ }\href@noop {} {\bibfield  {journal} {\bibinfo  {journal}
  {J. Phys. Soc. Jpn}\ }\textbf {\bibinfo {volume} {84}},\ \bibinfo {pages}
  {014704} (\bibinfo {year} {2015})}\BibitemShut {NoStop}%
\bibitem [{\citenamefont {Annett}\ and\ \citenamefont
  {Kruchinin}(2002)}]{Annett2002}%
  \BibitemOpen
  \bibinfo {editor} {\bibfnamefont {J.~F.}\ \bibnamefont {Annett}}\ and\
  \bibinfo {editor} {\bibfnamefont {S.}~\bibnamefont {Kruchinin}},\ eds.,\
  \href@noop {} {\emph {\bibinfo {title} {{NATO science Series New trends in
  superconductivity vol. 67}}}}\ (\bibinfo  {publisher} {Springer
  Netherlands},\ \bibinfo {year} {2002})\BibitemShut {NoStop}%
\bibitem [{\citenamefont {Hoogenboom}\ \emph {et~al.}(2003)\citenamefont
  {Hoogenboom}, \citenamefont {Berthod}, \citenamefont {Peter}, \citenamefont
  {Fischer}, ,\ and\ \citenamefont {Kordyuk}}]{Hoogenboom}%
  \BibitemOpen
  \bibfield  {author} {\bibinfo {author} {\bibfnamefont {B.~W.}\ \bibnamefont
  {Hoogenboom}}, \bibinfo {author} {\bibfnamefont {C.}~\bibnamefont {Berthod}},
  \bibinfo {author} {\bibfnamefont {M.}~\bibnamefont {Peter}}, \bibinfo
  {author} {\bibfnamefont {O.}~\bibnamefont {Fischer}}, , \ and\ \bibinfo
  {author} {\bibfnamefont {A.~A.}\ \bibnamefont {Kordyuk}},\ }\href@noop {}
  {\bibfield  {journal} {\bibinfo  {journal} {Phys. Rev. B.}\ }\textbf
  {\bibinfo {volume} {67}},\ \bibinfo {pages} {224502} (\bibinfo {year}
  {2003})}\BibitemShut {NoStop}%
\bibitem [{\citenamefont {Kordyuk}\ \emph {et~al.}(2003)\citenamefont
  {Kordyuk}, \citenamefont {Borisenko}, \citenamefont {Knupfer},\ and\
  \citenamefont {Fink}}]{Fink2003}%
  \BibitemOpen
  \bibfield  {author} {\bibinfo {author} {\bibfnamefont {A.~A.}\ \bibnamefont
  {Kordyuk}}, \bibinfo {author} {\bibfnamefont {S.~V.}\ \bibnamefont
  {Borisenko}}, \bibinfo {author} {\bibfnamefont {M.}~\bibnamefont {Knupfer}},
  \ and\ \bibinfo {author} {\bibfnamefont {J.}~\bibnamefont {Fink}},\
  }\href@noop {} {\bibfield  {journal} {\bibinfo  {journal} {Phys. Rev. B.}\
  }\textbf {\bibinfo {volume} {67}},\ \bibinfo {pages} {064504} (\bibinfo
  {year} {2003})}\BibitemShut {NoStop}%
\end{thebibliography}%
\end{document}